\PassOptionsToPackage{table}{xcolor}
\documentclass[sigconf,nonacm]{acmart}

\AtBeginDocument{%
  }

\setcopyright{acmlicensed}
\copyrightyear{2018}
\acmYear{2018}
\acmDOI{XXXXXXX.XXXXXXX}

\acmConference[Conference acronym 'XX]{Make sure to enter the correct
  conference title from your rights confirmation email}{June 03--05,
  2018}{Woodstock, NY}

\acmISBN{978-1-4503-XXXX-X/2018/06}
\usepackage{enumitem}
\usepackage{booktabs}
\usepackage{multirow}
\usepackage{makecell}
\usepackage{graphicx}
\usepackage[normalem]{ulem}
\usepackage{xcolor}
\definecolor{headerblue}{RGB}{228,238,249}
\definecolor{oursgreen}{RGB}{226,239,218}
\definecolor{datasetgray}{RGB}{242,242,242}

\usepackage{times}
\usepackage{latexsym}
\usepackage[T1]{fontenc}
\usepackage[utf8]{inputenc}
\usepackage{microtype}
\usepackage{graphicx}

\usepackage{booktabs}
\usepackage{multirow}
\usepackage{array}
\usepackage{graphicx}
\usepackage[table]{xcolor}
\usepackage{pifont}
\newcommand{\cmark}{\ding{51}}
\newcommand{\xmark}{\ding{55}}
\definecolor{oursblue}{RGB}{232,244,255}
\definecolor{groupgray}{RGB}{245,245,245}
\usepackage[most]{tcolorbox}
\usepackage{xcolor}
\usepackage{tabularx}
\definecolor{PromptBrown}{HTML}{7A5C3A}
\definecolor{PromptFrame}{HTML}{9A8468}
\definecolor{PromptBack}{HTML}{FAF8F4}

\newtcolorbox{promptbox}[1]{
    enhanced,
    breakable,
    colback=PromptBack,
    colframe=PromptFrame,
    coltitle=white,
    colbacktitle=PromptBrown,
    title={#1},
    fonttitle=\bfseries,
    boxrule=0.5pt,
    arc=1.8mm,
    outer arc=1.8mm,
    left=1.2mm,
    right=1.2mm,
    top=1mm,
    bottom=1mm,
    titlerule=0pt
}

\usepackage[most]{tcolorbox}
\usepackage{xcolor}

\definecolor{headerblue}{RGB}{232,240,248}
\definecolor{CaseBlue}{HTML}{3F5F73}
\definecolor{CaseFrame}{HTML}{8FA8B8}
\definecolor{CaseBack}{HTML}{F6FAFC}

\newtcolorbox{casebox}[1]{
    enhanced,
    breakable,
    fontupper=\small,
    colback=CaseBack,
    colframe=CaseFrame,
    coltitle=white,
    colbacktitle=CaseBlue,
    title={#1},
    fonttitle=\bfseries,
    boxrule=0.5pt,
    arc=1.8mm,
    outer arc=1.8mm,
    left=1.4mm,
    right=1.4mm,
    top=1mm,
    bottom=1mm,
    titlerule=0pt
}

\begin{document}

\title{AtomRec: Evolving Atomic Memory for Agentic Recommendation}

\author{%
Peiyu Hu\textsuperscript{1,2,*},
Weihai Lu\textsuperscript{3,*},
Siying Gu\textsuperscript{4, 2,*},
Zhuodong Liu\textsuperscript{5},
Zhaokai Luo\textsuperscript{2,\textdagger},
\\
Yuean Niu\textsuperscript{2}
Zhiyong Wang\textsuperscript{2},
Jia Wang\textsuperscript{1,\textdaggerdbl}
}

\affiliation{%
  \institution{%
      \textsuperscript{1}Xi'an Jiaotong-Liverpool University \quad
    \textsuperscript{2}Xiaohongshu \quad
    \textsuperscript{3}Peking University \quad
    \\
    \textsuperscript{4}East China Normal University
    \textsuperscript{5}Beijing Jiaotong University
  }
  \country{}
}

\email{%
  peiyuhu30@gmail.com,
  weihai.lu@pku.edu.cn,
  sy.gu@stu.ecnu.edu.cn,
  zhuodong.liu@bjtu.edu.cn,
  jia.wang02@xjtlu.edu.cn,
}

\email{%
{luozhaokai,niuyuean,sunzhenghuai}@xiaohongshu.com
}

\thanks{%
  \textsuperscript{*}These authors contributed equally to this work.\quad
  \textsuperscript{\textdagger}Team Leader.\quad
  \textsuperscript{\textdaggerdbl}Corresponding author: Jia Wang (jia.wang02@xjtlu.edu.cn).
}

\renewcommand{\shortauthors}{Hu et al.}


\begin{abstract}
Agentic recommender systems use large language models to maintain semantic memory and support evidence-aware recommendation. However, existing memory mechanisms often compress user and item information into coarse summaries and connect them with scalar collaborative links, making it difficult to preserve fine-grained preference stages or retrieve interpretable evidence as user interests evolve. We propose \textsc{AtomRec}, an agentic recommender with evolving atomic collaborative memory. \textsc{AtomRec} represents user and item memories as structured atomic units, builds semantic links across related memories, and evolves related historical fields when new interactions arrive. During recommendation, it retrieves linked memories as multi-hop evidence paths rather than isolated neighbor summaries, allowing collaborative signals to support grounded ranking. Experiments on four public benchmarks show that \textsc{AtomRec} consistently outperforms state-of-the-art agentic and memory-augmented baselines, with around 8.5\% average relative improvement across metrics.
\end{abstract}




\maketitle

\section{Introduction}

Recommender systems are moving from representation-based prediction toward agentic recommendation~\cite{rec1, rec2, rec3, agent4rec-survey1, agent4rec-survey2}. Traditional models encode user preferences with rating matrices, collaborative embeddings, or sequential hidden states~\cite{SASRec, sun2019bert4rec}. In contrast, LLM-powered recommender agents need to understand user intent, maintain long-term memory, integrate collaborative evidence, and support evidence-aware ranking~\cite{llm4rec-survey1, llm4rec-survey2, hu1, hu2, hu3}. Memory therefore becomes a key interface between historical interactions, collaborative signals, and language-based reasoning~\cite{memory-survey1, lu1, lu2, lu3}.

\begin{figure}[t]
    \centering
    \includegraphics[width=\linewidth]{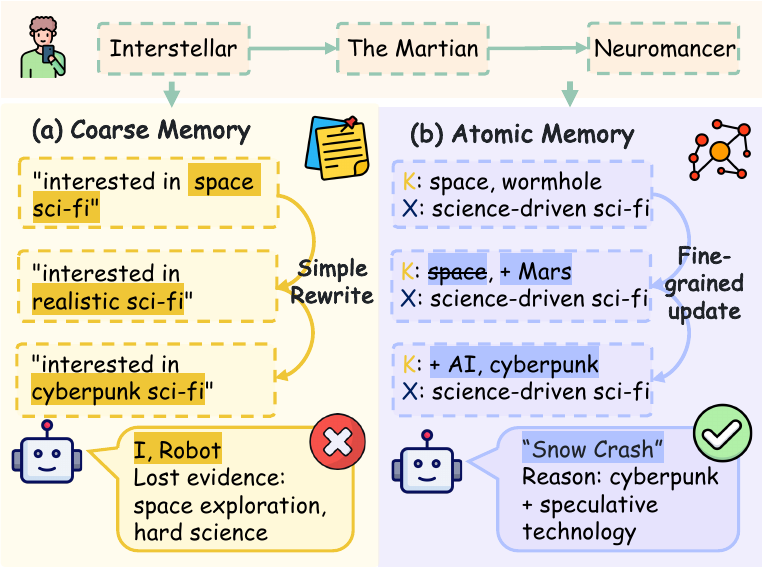}
    \caption{
    Motivation of \textsc{AtomRec}. Given the same evolving user history, coarse memory rewrites broad summaries and may lose fine-grained preference evidence, while atomic memory preserves preference traces as operational units for field/link evolution and evidence-path retrieval.
    }
    \label{fig:story_graph}
\end{figure}

In long-term recommendation, memory often needs to be fine-grained and dynamic~\cite{rec1, rec2, hu4, hu5}. A user's preference is not a single stable profile, but a collection of small and evolving traces, such as recurring genres, recent intents, item attributes, and collaborative signals from related users or items~\cite{rec1, rec2, lu4, lu5, lu6}. These traces may evolve at different speeds: some reflect stable long-term interests, while others capture emerging or context-dependent preferences. As illustrated in Figure~\ref{fig:story_graph}, a user may move from space exploration in \textit{Interstellar}, to realistic survival science fiction in \textit{The Martian}, and later to cyberpunk society in \textit{Neuromancer}. These stages remain related under the broad science-fiction theme, but each carries different recommendation evidence.

Recent agentic recommender systems use semantic memory to store user preferences and item properties, while collaborative memory, planner-based, and propagation-based agents further improve recommendation through user--item memory graphs, tool routing, information-gap analysis, or preference propagation~\cite{memrec, iagent, agentcf, chainrec, recthinker, recnet}. Despite this progress, existing memory mechanisms still face two key limitations\cite{lu7, zhang1, zhang2, zhang3, xy1, xy2}. \textbf{First, coarse granularity.} Summary-level memories compress multiple preference traces into one profile, making selective revision difficult. \textbf{Second, limited semantic operability.} Scalar interaction, similarity, or propagation weights capture relational strength, but provide little evidence about why memories are related~\cite{agent4rec-survey1, agent4rec-survey3, jgy1, jgy2}.

These limitations are especially harmful for long-term recommendation. As shown in Figure~\ref{fig:story_graph}, coarse updates may repeatedly rewrite a broad user profile, causing earlier fine-grained evidence, such as space exploration or hard-science survival, to be weakened by later interests. Meanwhile, retrieval often returns isolated neighbor summaries rather than evidence paths that explain how earlier preferences connect to emerging interests. As a result, the recommender agent may receive collaborative context, but still lack fine-grained and interpretable memory evidence for ranking.

To address the above limitations, inspired by recent atomic memory systems~\citep{a-mem,mem0}, we propose \textsc{AtomRec}, an LLM-based agentic recommender system with evolving atomic collaborative memory. The key idea is to make recommendation memory reorganizable rather than merely more detailed. \textsc{AtomRec} replaces coarse entity-level summaries with atomic memory units containing structured semantic fields, which serve as operational objects for locating, linking, revising, and preserving specific preference traces.

On top of these atomic units, \textsc{AtomRec} introduces Semantic Collaborative Link Construction and dynamic memory evolution. The linking module builds semantic relations across user and item atoms, turning scalar collaborative signals into interpretable evidence. The evolution module updates related historical atoms and their links when new interactions provide additional evidence, enabling the memory graph to reorganize over time instead of passively accumulating summaries. During recommendation, \textsc{AtomRec} retrieves along the evolved links to construct context-aware evidence paths for final ranking.

Our contributions are as follows:
\begin{enumerate}
    \item We propose \textsc{AtomRec}, an agentic recommender system that represents user and item memories as linked and evolving atomic units.
    \item We introduce Semantic Collaborative Link Construction and dynamic memory evolution to support fine-grained semantic reorganization and context-aware evidence retrieval.
    \item We conduct experiments on four public benchmarks, showing that \textsc{AtomRec} outperforms strong baselines with around 8.5\% average relative improvement and better handles users with stronger preference drift.
\end{enumerate}

\section{Method}
\subsection{Problem Formulation}

Let $\mathcal{U}$ and $\mathcal{I}$ denote the user and item sets. For a target user $u$ at time $t$, we denote the historical interactions as $\mathcal{H}_u^{t-1}$ and the candidate set as $\mathcal{C}_u^t \subseteq \mathcal{I}$. Given a natural language instruction $I_u$, the task is to rank candidates according to the user's current preference.

\textsc{AtomRec} maintains a memory space $\mathcal{M}$ of user and item memories, where each memory is an evolvable atomic note and semantic links connect related notes. Let $M_u^t$ denote the pre-$t$ user memory, $M_i^t$ the memory of candidate item $i$, and $M_{\mathrm{collab}}^t$ the collaborative evidence retrieved from $\mathcal{M}$. The recommendation agent scores each candidate by
\[
r_{u,i}^{t} =
\mathrm{LM}_{\mathrm{Rec}}
\left(
I_u, M_u^t, M_{\mathrm{collab}}^t, M_i^t
\right),
\quad i \in \mathcal{C}_u^t .
\]
Candidates are ranked by $r_{u,i}^{t}$. All memory construction, linking, evolution, and retrieval use only interactions before time $t$; the target item is used only as a candidate during final reranking and is never written back into memory for the same test instance.

\subsection{Overview}

Recommendation memory contains heterogeneous preference traces, such as stable interests, recent intents, item attributes, and collaborative signals. Existing memory-based and agentic recommenders often compress these traces into coarse user/item summaries and connect them with interaction, similarity, or propagation weights~\citep{memrec, iagent, recbot, agentcf, agent4rec-survey1, agent4rec-survey3}. This can obscure preference stages and provide limited semantic evidence for selective revision and evidence-aware retrieval.

\begin{figure*}[t]
    \centering
    \includegraphics[width=0.99\linewidth]{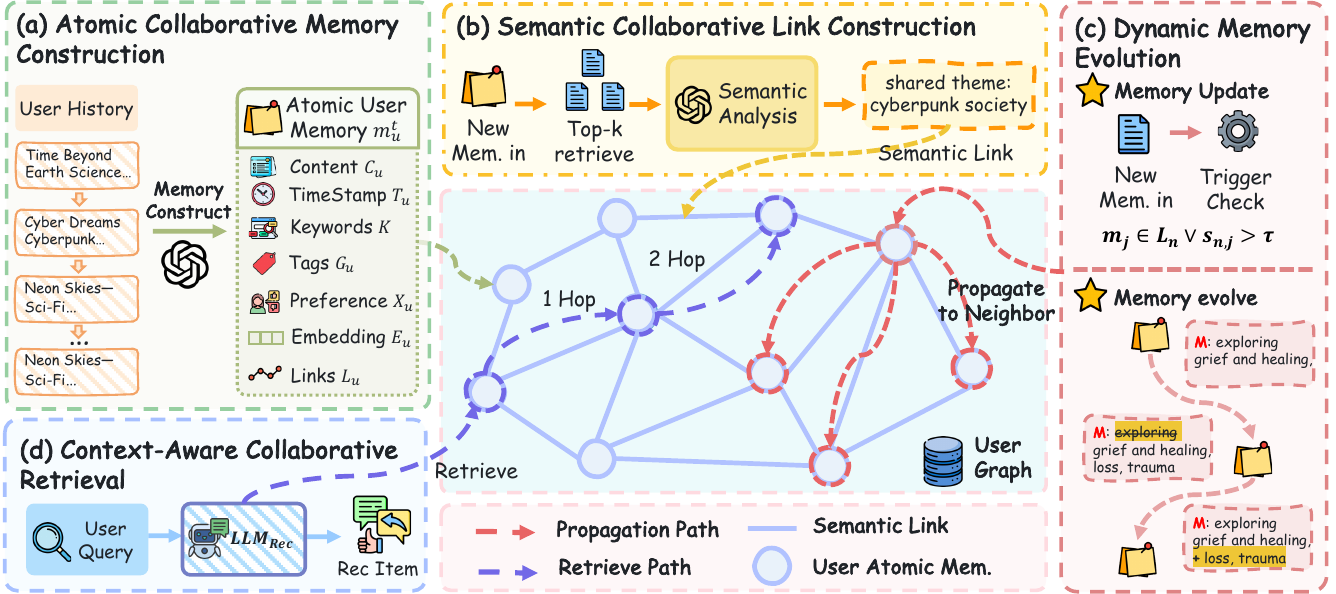}
    \caption{
    Architecture of \textsc{AtomRec}. Atomic memory construction, semantic collaborative link construction, dynamic memory evolution, and context-aware collaborative retrieval jointly support evidence-path recommendation.
    }
    \label{fig:arc}
\end{figure*}

To address these limitations, we propose \textsc{AtomRec}, an agentic recommender with evolving atomic collaborative memory. As shown in Figure~\ref{fig:arc}, \textsc{AtomRec} follows four steps. First, \textbf{atomic collaborative memory construction} represents user and item memories as structured notes, providing fine-grained units for selective retrieval and revision. Second, \textbf{Semantic Collaborative Link Construction} converts nearby memories into semantic relations, making collaborative links explainable rather than only numerical. Third, \textbf{dynamic memory evolution} updates related historical notes when new evidence arrives, allowing memory to track emerging interests while preserving temporal traces. Finally, \textbf{context-aware collaborative retrieval} expands along semantic links and synthesizes multi-hop evidence paths for grounded ranking.

\subsection{Atomic Collaborative Memory Construction}

\textsc{AtomRec} represents each user or item memory as an atomic note rather than a single coarse summary. For a user $u$ at time $t$, the atomic memory note is defined as
\[
m_u^t = \{c_u^t, t_u, K_u^t, G_u^t, X_u^t, e_u^t, L_u^t\},
\]
where $c_u^t$ denotes the textual memory content, $t_u$ is the timestamp, $K_u^t$ is a set of keywords, $G_u^t$ is a set of semantic tags, $X_u^t$ is a short contextual description, $e_u^t$ is a dense embedding, and $L_u^t$ stores linked memory identifiers. Item memories follow the same structure. The embedding of each memory note is computed from its semantic fields:
\[
e_u^t =
f_{\mathrm{enc}}
\left(
\operatorname{concat}
(c_u^t, K_u^t, G_u^t, X_u^t)
\right),
\]
where $f_{\mathrm{enc}}$ is a text encoder.

This structured note design makes memory both retrievable and interpretable: the content field preserves original preference evidence, keywords and tags expose explicit semantic facets, the context field summarizes current intent, and the link field allows the note to participate in the collaborative memory graph. We do not assume that these fields are fully disentangled; instead, we use them as complementary semantic views that make retrieval, linking, and evolution more controllable. Compared with coarse user profiles, atomic notes reduce the tendency to collapse heterogeneous interests into a single summary and provide a finer-grained substrate for selectively revising or retrieving preference evidence.

\subsection{Semantic Collaborative Link Construction}

Traditional collaborative memory usually connects users and items through predefined interaction edges, co-occurrence statistics, or scalar similarity weights. These edges indicate collaborative strength, but they do not explain the semantic relation between memories. \textsc{AtomRec} instead builds semantic links among atomic memory notes.

When a new memory note $m_n$ is created, we first retrieve semantically nearby memories from the memory space $\mathcal{M}$ using embedding similarity:

\[
s_{n,j}
=
\frac{e_n^\top e_j}{\|e_n\|\|e_j\|},
\quad
m_j \in \mathcal{M}.
\]

The top-$k$ memories form the candidate set:

\[
\mathcal{M}_{\mathrm{near}}^n
=
\operatorname{TopK}_{m_j \in \mathcal{M}}(s_{n,j}).
\]

Then, the memory agent $\mathrm{LM}_{\mathrm{Mem}}$ analyzes the new memory and its candidate memories to generate semantic links:

\[
L_n
\leftarrow
\mathrm{LM}_{\mathrm{Mem}}
\left(
m_n \parallel \mathcal{M}_{\mathrm{near}}^n \parallel P_{\mathrm{link}}
\right),
\]

where $P_{\mathrm{link}}$ is the linking prompt. It guides $\mathrm{LM}_{\mathrm{Mem}}$ to identify semantic relations between memory notes, including shared topics, complementary preferences, intent progression, preference-transition relations, and cross-domain transfer. The output $L_n$ is a set of linked memory identifiers, optionally with relation descriptions, which is stored in the link field of $m_n$. In this way, collaborative relations become semantic and interpretable, rather than only numerical edge weights.

\subsection{Dynamic Memory Evolution}

A new interaction may change the interpretation of previous memories. Instead of only appending new memories, \textsc{AtomRec} allows new evidence to trigger updates to related historical memories. For each candidate memory $m_j \in \mathcal{M}_{\mathrm{near}}^n$, we decide whether to evolve it using a simple trigger condition:

\[
\text{Evolve}(m_j)
=
\mathbb{I}
\left[
m_j \in L_n
\ \lor\
s_{n,j}>\tau_{\mathrm{evo}}
\right],
\]
where $L_n$ is the semantic link set of the new memory, $s_{n,j}$ is the similarity score between $m_n$ and $m_j$, and $\tau_{\mathrm{evo}}$ is the memory evolution threshold. If the trigger condition is satisfied, the memory agent updates the historical memory:

\[
m_j^{*}
\leftarrow
\mathrm{LM}_{\mathrm{Mem}}
\left(
m_n
\parallel
m_j
\parallel
\mathcal{M}_{\mathrm{near}}^n
\parallel
P_{\mathrm{evolve}}
\right),
\]

where $P_{\mathrm{evolve}}$ instructs the agent to refine the memory while preserving its original semantics. The update is performed at the field level:

\[
m_j^{*}
=
\{c_j^{*}, t_j, K_j^{*}, G_j^{*}, X_j^{*}, e_j^{*}, L_j^{*}\}.
\]

Specifically, evolution may strengthen keywords, refine semantic tags, reconstruct the contextual description, and update semantic links. The timestamp $t_j$ remains unchanged to preserve the original temporal position, while the semantic fields are revised to reflect newly observed evidence. This field-level evolution allows the memory graph to reorganize itself as user behavior changes.

\subsection{Context-Aware Collaborative Retrieval and Recommendation}

During recommendation, \textsc{AtomRec} retrieves a linked memory subgraph instead of isolated top-$k$ neighbors. Given the current user state, we first retrieve an initial set of relevant memories:

\[
\mathcal{M}_{0}^{t}
=
\operatorname{TopK}_{m_i \in \mathcal{M}}
\left(
\cos(q_u^t, e_i)
\right),
\]

where $q_u^t$ is the query representation derived from the current instruction, the user's recent interactions, and the pre-$t$ atomic memory fields using the same encoder $f_{\mathrm{enc}}$. We then expand the retrieved memories through semantic links:

\[
\mathcal{G}_{\mathrm{sub}}^{t}
=
\mathcal{M}_{0}^{t}
\cup
\bigcup_{m_i \in \mathcal{M}_{0}^{t}} L_i,
\]

where the expansion can be extended to multiple hops in practice. We use relation descriptions as textual evidence for synthesis, while hop depth and retrieval similarity control subgraph expansion. The resulting subgraph contains not only similar memories, but also semantically linked evidence from related users and items.

The memory agent summarizes this subgraph into collaborative evidence:

\[
M_{\mathrm{collab}}^t
\leftarrow
\mathrm{LM}_{\mathrm{Mem}}
\left(
\mathcal{G}_{\mathrm{sub}}^{t}
\parallel
P_{\mathrm{synth}}
\right),
\]

where $P_{\mathrm{synth}}$ guides the agent to extract preference facets and evidence paths. Finally, the recommendation agent scores each candidate item $i \in \mathcal{C}_u^t$ based on the user's memory, collaborative evidence, and item memory:

\[
r_{u,i}^{t}
=
\mathrm{LM}_{\mathrm{Rec}}
\left(
M_u^t
\parallel
M_{\mathrm{collab}}^t
\parallel
M_i^t
\parallel
P_{\mathrm{rank}}
\right).
\]

Candidate items are ranked by $r_{u,i}^{t}$. Because the retrieved context is built from semantic links among atomic memories, the recommendation agent receives an evidence path rather than a set of disconnected neighbor summaries. This enables \textsc{AtomRec} to combine semantic reasoning, collaborative enhancement, and memory evolution in a unified recommendation process.

\section{Experimental Setup}

We conduct experiments to answer the following research questions:
\begin{itemize}
    \setlength{\itemsep}{2pt}
    \setlength{\parsep}{0pt}

    \item \textbf{RQ1 (Overall Performance):} Does \textsc{AtomRec} consistently improve recommendation accuracy over traditional, LM-based, agentic, and memory-augmented baselines?

    \item \textbf{RQ2 (Component Analysis):} How do atomic memory construction, semantic linking, memory evolution, link-aware retrieval, and key hyperparameters affect performance?

    \item \textbf{RQ3 (Preference Drift):} Does \textsc{AtomRec} better capture users whose preferences shift over time?

    \item \textbf{RQ4 (Qualitative Behavior):} How does evolving atomic memory support interpretable evidence-path retrieval in concrete cases?

    \item \textbf{RQ5 (Efficiency):} What cost-performance trade-offs does \textsc{AtomRec} exhibit under different methods and backbone models?
\end{itemize}

\textbf{Datasets.} We evaluate on four instruction-augmented benchmarks from MemRec~\citep{memrec}: Amazon Books, Goodreads, MovieTV, and Yelp, originally adapted from InstructRec~\citep{instructRec}. We use the same preprocessed data and train/dev/test splits for fair comparison~\citep{SASRec, amazon-dataset}. Dataset statistics are provided in Appendix~\ref{app:dataset_details}.

\textbf{Baselines.} We compare \textsc{AtomRec} with representative baselines from three categories: traditional recommenders, including LightGCN~\citep{lightgcn} and SASRec~\citep{SASRec}; LM-based recommenders, including P5~\citep{p5} and Vanilla LLM~\citep{vanilla}; and agentic or memory-augmented recommenders, including iAgent~\citep{iagent}, RecBot~\citep{recbot}, AgentCF~\citep{agentcf}, i2Agent~\citep{iagent}, and MemRec~\citep{memrec}. Detailed baseline descriptions are provided in Appendix~\ref{app:baseline_details}.

\textbf{Evaluation Protocols.} We evaluate the main results on the full test sets with candidate size $N=10$, reporting H@K and N@K for $K \in \{3,5\}$. For ablation and analysis experiments, we use the same randomly sampled 1,000 users across methods following prior agentic recommendation settings~\cite{iagent, memrec}.

\textbf{Implementation Details.} We use \texttt{gpt-4o-mini}~\citep{gpt4o} as both the memory agent $\mathrm{LM}_{\mathrm{Mem}}$ and the recommendation agent $\mathrm{LM}_{\mathrm{Rec}}$ in the main experiments. We use frozen Sentence-T5~\citep{sentence-t5} as the text encoder $f_{\mathrm{enc}}$ to encode concatenated atomic fields for cosine-similarity retrieval. Unless otherwise specified, we set $k_{\mathrm{link}}=20$, $\tau_{\mathrm{evo}}=0.7$, and $h=2$; these values are selected on the validation set and examined in the hyperparameter sensitivity analysis in Section~\ref{sec:hyperparameter_sensitivity}. The memory agent constructs atomic memories, generates semantic links, evolves related historical memories, and synthesizes collaborative evidence, while the recommendation agent ranks candidate items based on the user instruction, user memory, collaborative evidence, and candidate item memories. During evaluation, memory states are built only from pre-target interactions, and the target item is used only for final reranking. More implementation details are provided in Appendix~\ref{app:implementation_details}.

\section{Experimental Results}

\begin{table*}[t]
\centering
\small
\setlength{\tabcolsep}{3.6pt}
\renewcommand{\arraystretch}{1.08}
\caption{
Main results on four datasets. ``Improv.'' denotes the relative improvement of our method over the best baseline. The best results are highlighted in bold, and the second-best results are underlined. All improvements over the strongest baseline are significant under paired bootstrap testing ($p<0.05$).
}
\label{tab:main_results}

\resizebox{\textwidth}{!}{
\begin{tabular}{
l
@{\hskip 6pt}cccc
@{\hskip 6pt}cccc
@{\hskip 6pt}cccc
@{\hskip 6pt}cccc
}
\toprule
\multirow{2}{*}{Model}
& \multicolumn{4}{c}{Books}
& \multicolumn{4}{c}{Goodreads}
& \multicolumn{4}{c}{MovieTV}
& \multicolumn{4}{c}{Yelp} \\
\cmidrule(lr){2-5}
\cmidrule(lr){6-9}
\cmidrule(lr){10-13}
\cmidrule(lr){14-17}
& H@3 & N@3 & H@5 & N@5
& H@3 & N@3 & H@5 & N@5
& H@3 & N@3 & H@5 & N@5
& H@3 & N@3 & H@5 & N@5 \\
\midrule

\rowcolor{groupgray}
\multicolumn{17}{l}{\textit{Traditional Recommenders}} \\

LightGCN
& 0.3259 & 0.2596 & 0.5703 & 0.3592
& 0.5879 & 0.4432 & 0.7903 & 0.5263
& 0.5643 & 0.4738 & 0.6883 & 0.5241
& 0.5658 & 0.4720 & 0.7546 & 0.5494 \\

SASRec
& 0.2830 & 0.2001 & 0.4845 & 0.2824
& 0.3518 & 0.2576 & 0.5407 & 0.3349
& 0.5233 & 0.4470 & 0.6382 & 0.4942
& 0.4312 & 0.3458 & 0.5597 & 0.3980 \\

\midrule
\rowcolor{groupgray}
\multicolumn{17}{l}{\textit{LM-based Recommenders}} \\

P5
& 0.3607 & 0.2994 & 0.5273 & 0.3671
& 0.3229 & 0.2509 & 0.5060 & 0.3256
& 0.3206 & 0.2554 & 0.5008 & 0.3290
& 0.3207 & 0.2435 & 0.5220 & 0.4785 \\

Vanilla LLM
& 0.5617 & 0.4533 & 0.7270 & 0.5226
& 0.4662 & 0.3948 & 0.7390 & 0.5041
& 0.7564 & 0.6098 & 0.8603 & 0.6445
& 0.5275 & 0.3696 & 0.6861 & 0.4360 \\

\midrule
\rowcolor{groupgray}
\multicolumn{17}{l}{\textit{Agentic Recommenders}} \\

iAgent
& 0.5560 & 0.4858 & 0.6905 & 0.5409
& 0.4949 & 0.3954 & 0.6591 & 0.4626
& 0.6170 & 0.5361 & 0.7420 & 0.5871
& 0.6005 & 0.5148 & 0.7300 & 0.5681 \\

RecBot
& 0.5491 & 0.4846 & 0.6786 & 0.5376
& 0.4754 & 0.3876 & 0.6495 & 0.4589
& 0.6113 & 0.5375 & 0.7309 & 0.5866
& 0.6003 & 0.5156 & 0.7169 & 0.5636 \\

AgentCF
& 0.6060 & 0.4960 & 0.7403 & 0.5512
& 0.5910 & 0.4654 & 0.7726 & 0.5399
& 0.6693 & 0.5523 & 0.7864 & 0.6006
& 0.4374 & 0.3326 & 0.6374 & 0.4147 \\

i2Agent
& 0.6517 & 0.5649 & 0.7708 & 0.6138
& 0.6079 & 0.4825 & 0.7675 & 0.5481
& 0.7225 & 0.6262 & 0.8221 & 0.6672
& 0.6454 & 0.5517 & 0.7648 & 0.6007 \\

MemRec
& \underline{0.6786} & \underline{0.6078} & \underline{0.7764} & \underline{0.6480}
& \underline{0.6498} & \underline{0.5490} & \underline{0.7991} & \underline{0.6005}
& \underline{0.7660} & \underline{0.6907} & \underline{0.8654} & \underline{0.7068}
& \underline{0.6632} & \underline{0.5914} & \underline{0.7738} & \underline{0.6251} \\

\midrule
\rowcolor{oursblue}
\textbf{Ours}
& \textbf{0.7462} & \textbf{0.6690} & \textbf{0.8543} & \textbf{0.7130}
& \textbf{0.6980} & \textbf{0.5922} & \textbf{0.8558} & \textbf{0.6458}
& \textbf{0.8323} & \textbf{0.7532} & \textbf{0.9364} & \textbf{0.7683}
& \textbf{0.7119} & \textbf{0.6393} & \textbf{0.8330} & \textbf{0.6745} \\

\midrule
Improv.
& 9.96\% & 10.07\% & 10.03\% & 10.03\%
& 7.42\% & 7.86\% & 7.10\% & 7.55\%
& 8.65\% & 9.05\% & 8.20\% & 8.70\%
& 7.35\% & 8.10\% & 7.65\% & 7.90\% \\

\bottomrule
\end{tabular}
}
\end{table*}

\subsection{Performance (RQ1)}

Table~\ref{tab:main_results} presents the overall comparison across four datasets. \textsc{AtomRec} achieves the best results on all datasets and all metrics, improving over the strongest baseline by around 8.5\% across metrics.

\begin{itemize}
    \setlength{\itemsep}{2pt}
    \setlength{\parsep}{0pt}

    \item \textsc{AtomRec} consistently outperforms all baselines on Books, Goodreads, MovieTV, and Yelp. The larger gains on Books and MovieTV suggest that atomic collaborative memory is especially useful for sparse and content-driven recommendation, showing \textbf{consistent cross-domain gains}.

    \item Agentic and memory-augmented baselines generally outperform traditional and LM-only methods, indicating that explicit semantic memory and collaborative evidence are important for instruction-aware ranking, confirming \textbf{the value of agentic memory}.

    \item Compared with MemRec, \textsc{AtomRec} further benefits from structured fields, semantic links, and field-level evolution. These designs enable connected evidence-path retrieval rather than isolated neighbor summaries, supporting \textbf{fine-grained atomic evidence}.
\end{itemize}

\subsection{Ablation Study (RQ2)}

Table~\ref{tab:ablation_books} reports the ablation results on Books, with additional datasets in Appendix~\ref{app:additional_ablation}. Removing any component hurts performance, confirming that \textsc{AtomRec}'s gains come from the joint design of atomic memory, semantic linking, memory evolution, and link-aware retrieval.

\begin{itemize}
    \setlength{\itemsep}{2pt}
    \setlength{\parsep}{0pt}
    \item \textbf{Atomic fields matter.} Collapsing atomic notes into coarse summaries causes the largest drop, while removing keywords/tags or context also hurts performance.
    \item \textbf{Semantic links help.} Removing links or replacing them with embedding-only neighbors reduces performance, showing the value of relation-aware evidence.
    \item \textbf{Evolution and path retrieval help.} Append-only memory and removing link-aware retrieval both underperform, showing that evolving memories and connected evidence paths improve ranking.
\end{itemize}

\begin{table}[t]
\centering
\normalsize
\setlength{\tabcolsep}{4.5pt}
\renewcommand{\arraystretch}{1.15}
\caption{
Ablation study of \textsc{AtomRec} on the Books dataset. Avg. $\Delta$ denotes the average relative performance decrease across all four metrics compared with the full model.
}
\label{tab:ablation_books}
\resizebox{\linewidth}{!}{
\begin{tabular}{lccccc}
\toprule
Variant & H@3 & N@3 & H@5 & N@5 & $\Delta_{\mathrm{avg}}$ \\
\midrule
\rowcolor{oursblue}
Full \textsc{AtomRec} 
& \textbf{0.7462} & \textbf{0.6690} & \textbf{0.8543} & \textbf{0.7130} & -- \\

\midrule
w/o Atomic Mem. 
& 0.7046 & 0.6308 & 0.8109 & 0.6742 & 5.45\% \\

w/o Keywords/Tags
& 0.7159 & 0.6416 & 0.8238 & 0.6854 & 3.90\% \\

w/o Context Field
& 0.7227 & 0.6480 & 0.8320 & 0.6927 & 2.94\% \\

\midrule
w/o Collab. Link. 
& 0.7168 & 0.6424 & 0.8255 & 0.6860 & 3.77\% \\

Embedding-only Links
& 0.7235 & 0.6496 & 0.8334 & 0.6936 & 2.78\% \\

\midrule
Append-only Memory
& 0.7251 & 0.6512 & 0.8350 & 0.6951 & 2.56\% \\

w/o Link Retrieval 
& 0.7292 & 0.6567 & 0.8391 & 0.7004 & 1.92\% \\

\bottomrule
\end{tabular}
}
\end{table}

\subsection{Hyperparameter Sensitivity (RQ2)}
\label{sec:hyperparameter_sensitivity}

\begin{figure}[t]
    \centering
    \includegraphics[width=0.98\columnwidth]{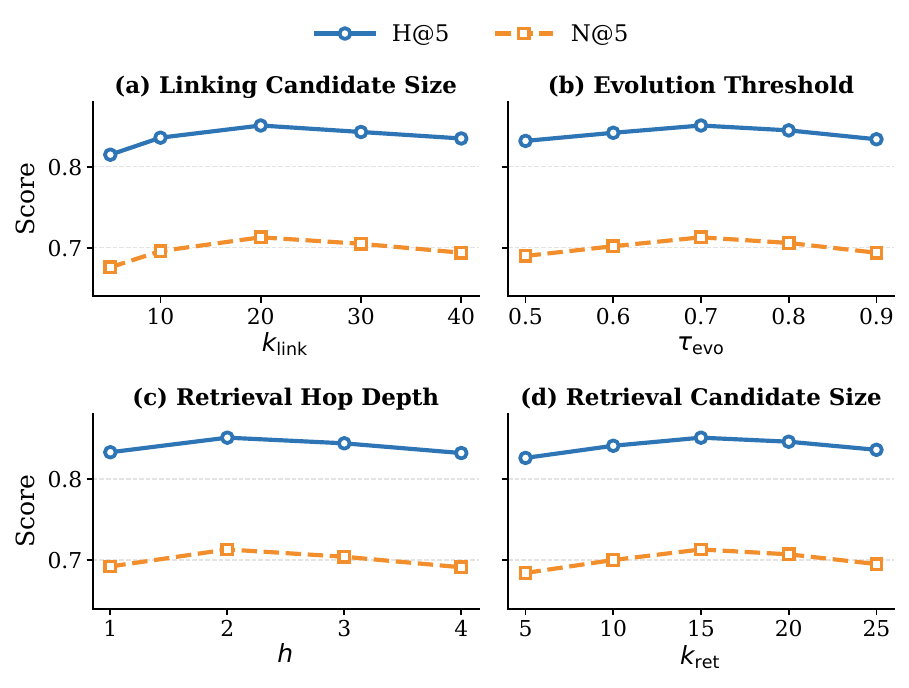}
    \caption{
   Hyperparameter sensitivity of \textsc{AtomRec} on Books. We report H@5 and N@5 while varying the linking candidate size $k_{\mathrm{link}}$, memory evolution threshold $\tau_{\mathrm{evo}}$, retrieval hop depth $h$, and retrieval candidate size $k_{\mathrm{ret}}$.
    }
    \label{fig:hyper_sensitivity}
\end{figure}

Figure~\ref{fig:hyper_sensitivity} shows the sensitivity of four memory-side hyperparameters on Books using H@5 and N@5. \textsc{AtomRec} performs best when $k_{\mathrm{link}}=20$, $\tau_{\mathrm{evo}}=0.7$, $h=2$, and $k_{\mathrm{ret}}=15$. These results suggest that the memory mechanism benefits from moderate semantic linking, conservative memory evolution, limited multi-hop retrieval, and a compact retrieval candidate set. Smaller values may miss useful collaborative evidence, while larger values can introduce noisy links or weakly related memories.

\subsection{Preference Drift Analysis (RQ3)}

\begin{figure}[t]
    \centering
    \includegraphics[width=0.96\columnwidth]{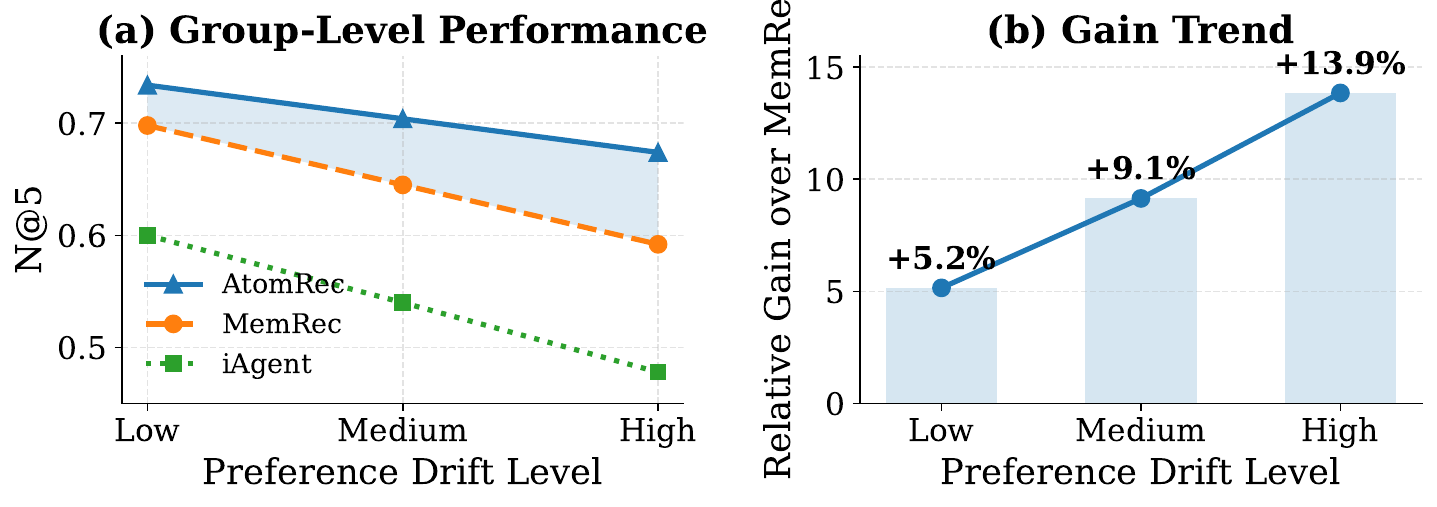}
    \caption{
    Preference drift analysis.
    (a) N@5 performance across user groups with different drift levels.
    (b) Relative gain of \textsc{AtomRec} over MemRec increases as preference drift becomes stronger.
    }
    \label{fig:drift_analysis}
\end{figure}

We examine whether \textsc{AtomRec} better handles users with changing interests. For each user, we split the history into early and late segments and compute the drift score as $d_u = 1 - \cos(\bar{e}_u^{\mathrm{early}}, \bar{e}_u^{\mathrm{late}})$, where $\bar{e}_u^{\mathrm{early}}$ and $\bar{e}_u^{\mathrm{late}}$ are the averaged item embeddings of the two segments. Users are then divided into Low, Medium, and High Drift groups. As shown in Figure~\ref{fig:drift_analysis}, all methods degrade as preference drift increases, while \textsc{AtomRec} obtains larger gains over MemRec on higher-drift users. This suggests that field-level memory evolution and semantic evidence paths help preserve emerging interests under preference shifts.

\subsection{Case Study (RQ4)}

\begin{figure*}[t]
    \centering
    \includegraphics[width=0.99\textwidth]{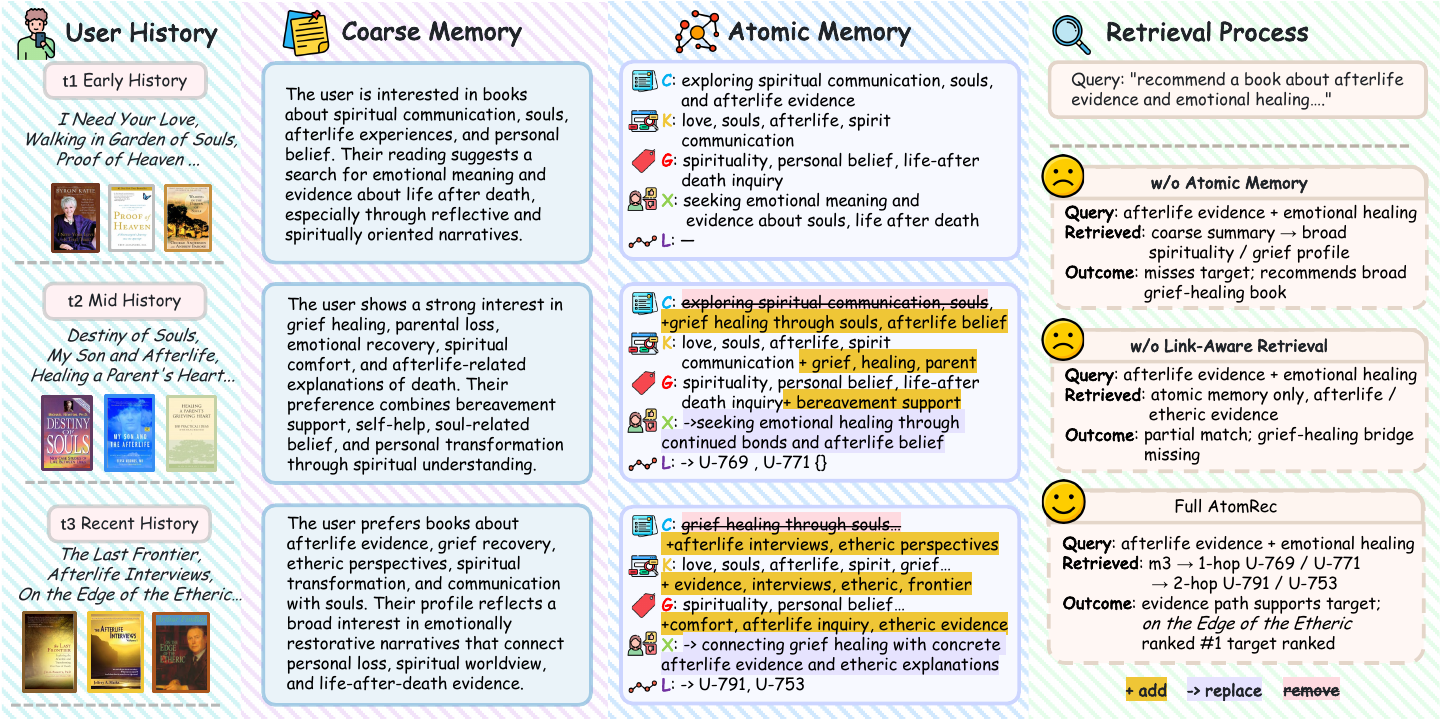}
    \caption{
    Case study of coarse memory rewriting and atomic memory evolution on Amazon Books. \textsc{AtomRec} preserves structured atomic fields and uses link-aware retrieval to recover the target item.
    }
    \label{fig:case_study}
\end{figure*}

Figure~\ref{fig:case_study} shows a representative case where the user's history shifts from spiritual communication, to grief healing, and then to etheric and afterlife-evidence books. Coarse memory captures these broad themes but repeatedly rewrites them into one summary, mixing preference stages and weakening stage-level evidence. In contrast, \textsc{AtomRec} preserves atomic fields and performs field-level edits, keeping earlier evidence accessible while refining current intent. The retrieval comparison further shows that the target item, \textit{On the Edge of the Etheric}, is missed without atomic memory and only partially matched without link-aware retrieval, while full \textsc{AtomRec} follows 1-hop and 2-hop evidence links to rank the target item first.

\subsection{Efficiency and Deployment Analysis (RQ5)}
\label{sec:efficiency_cost}

\begin{figure}[t]
    \centering
    \includegraphics[width=0.99\columnwidth]{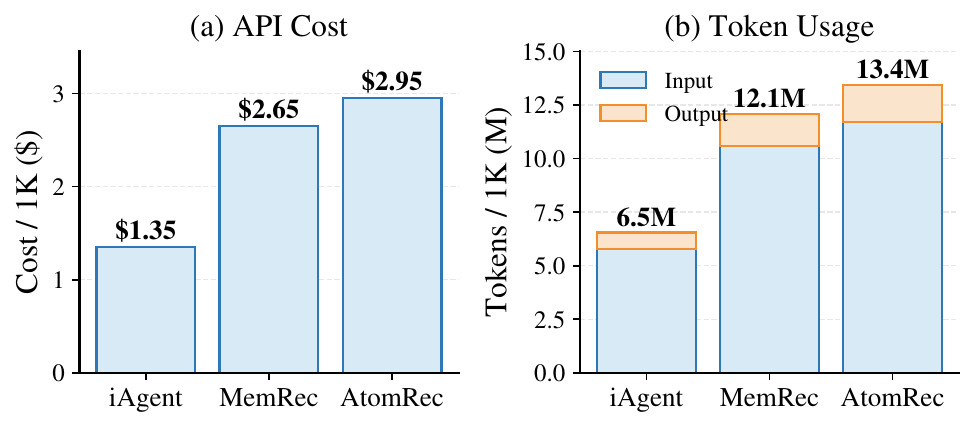}
    \caption{
    Absolute efficiency breakdown under the default lightweight backbone. We report estimated API cost and token usage per 1K interactions.
    }
    \label{fig:efficiency_breakdown_main}
\end{figure}

We analyze deployment cost using cost-performance trends and token/cost statistics, with details in Appendix~\ref{app:efficiency_details}. Figures~\ref{fig:cost_tradeoff} and~\ref{fig:efficiency_breakdown_main} show two findings:

\begin{itemize}
    \setlength{\itemsep}{2pt}
    \setlength{\parsep}{0pt}

    \item \textbf{Cost-performance trade-off.} \textsc{AtomRec} achieves higher N@5 on Books than representative LLM-based and agentic baselines with moderate extra cost. Stronger backbones bring only modest additional gains, suggesting that the improvement mainly comes from the memory mechanism rather than backbone scaling alone.

    \item \textbf{Online reranking overhead.} \textsc{AtomRec} uses more tokens than MemRec due to atomic construction, semantic linking, field-level evolution, and evidence synthesis. Since these memory-side operations can be cached and executed asynchronously, the additional online reranking overhead remains limited.
\end{itemize}

\begin{figure}[t]
    \centering
    \includegraphics[width=0.99\columnwidth]{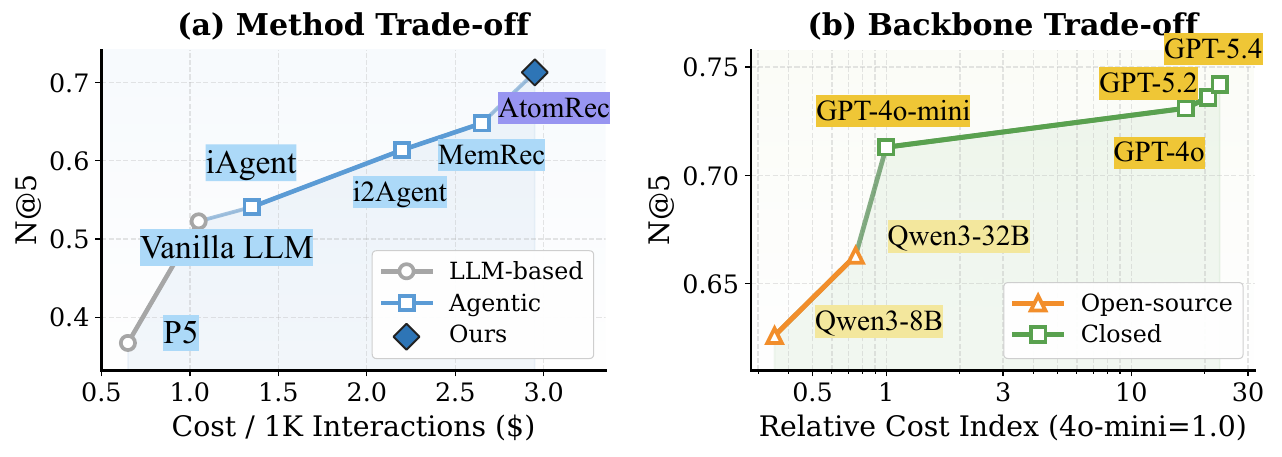}
   \caption{ Efficiency-aware analysis of \textsc{AtomRec} on Books. (a) Method-level cost-performance trade-off. (b) Backbone-level cost-performance trade-off, with cost normalized by gpt-4o-mini.
    }
    \label{fig:cost_tradeoff}
\end{figure}

\section{Related Work}

\textbf{Memory in LLM Agents.} While LLMs excel in long-horizon reasoning \cite{gpt4, gpt-hf, gpt-zeroshot, react, zj1, zj2, alan1, alan2}, their restricted context windows necessitate external memory for persistent knowledge retention \cite{memory-survey1, memory-survey2, alan3, alan4, alan5}. Early architectures rely on predefined workflows, like MemoryBank's forgetting curves \cite{memorybank}, MemGPT's hierarchical buffers \cite{memgpt}, and SCM's read-write streams \cite{scm}. Later models (e.g., A-Mem \cite{a-mem}, Mem0 \cite{mem0}, MemInsight \cite{meminsight}) enhance adaptability via associative linking and layered summarization \cite{selfevolve, rag-survey}. However, they predominantly focus on single-agent memory, leaving collaborative memory across users and items underexplored for interactive recommendation.

\textbf{LLM-based Recommendation Agents.} LLM agents support recommendation through simulation, planning, tool use, and memory-driven personalization~\cite{agent4rec-survey1, arvin1, arvin2, arvin3, arvin4, liang2026learn}. Agent4Rec~\cite{agent4rec} and AgentCF~\cite{agentcf} simulate user--item dynamics, while RecMind~\cite{recmind}, InteRecAgent~\cite{interrec}, and MACRec~\cite{macrec} explore task-oriented agent workflows. Recent systems further improve adaptive reasoning: ChainRec~\cite{chainrec} routes standardized tools, RecThinker~\cite{recthinker} plans tool calls through information-gap analysis, and RecNet~\cite{recnet} propagates preference updates through router agents. For long-term personalization, iAgent~\cite{iagent} and RecBot~\cite{recbot} update isolated user profiles, while MemRec~\cite{memrec} builds a collaborative memory graph. In contrast, \textsc{AtomRec} focuses on the memory substrate itself by decomposing coarse memories into field-structured atomic notes, linking them semantically, and evolving related historical memories for evidence-path recommendation.

\section{Conclusion}

We present \textsc{AtomRec}, an agentic recommender system with evolving atomic collaborative memory. \textsc{AtomRec} represents user and item memories as structured atomic notes, connects them through semantic links, and evolves related historical memories when new interactions arrive. By replacing coarse memory summaries and scalar collaborative edges with atomic memory units and semantic evidence paths, \textsc{AtomRec} supports fine-grained long-term preference modeling. Experiments on four instruction-augmented benchmarks show consistent improvements over traditional, LLM-based, agentic, and memory-augmented baselines, while ablation studies and qualitative analysis confirm the benefits of atomic construction, semantic collaborative link construction, and dynamic evolution. These results suggest that making memory more structured, linkable, and evolvable is a promising direction for building more interpretable and adaptive recommender agents.

\section*{Limitations}

\textsc{AtomRec} has three main limitations. First, memory evolution may over-compress user interests into a dominant subtheme, causing semantically adjacent false positives when fine-grained candidate discrimination is required. Our qualitative analysis in Appendix~\ref{app:qualitative} shows that the model can capture the correct broad preference region but still under-rank the target item when nearby candidates share partial emotional or spiritual cues. Second, the model may over-emphasize stable long-term preference trajectories and under-rank idiosyncratic or short-term exploratory targets. Third, collaborative signals may be absorbed into memory fields without being preserved as explicit final-step links, which weakens post-hoc traceability. Future work should improve fine-grained reranking, uncertainty-aware modeling of exploratory behavior, and provenance preservation for semantic memory links.

\bibliographystyle{ACM-Reference-Format}
\bibliography{refs/refs}

\appendix
\clearpage

\section{Dataset Details}
\label{app:dataset_details}

We used the same preprocessed representation and train/dev/test splits as MemRec~\cite{memrec} for all datasets. The four benchmarks were originally adapted from InstructRec~\cite{instructRec} and cover different recommendation domains, including book recommendation, social reading, movies and TV shows, and local business services. Table~\ref{tab:dataset_stats_app} reports the dataset statistics used in our experiments.

\textbf{Books.}~\cite{amazon-dataset, amazon-dataset1} is a large-scale book recommendation benchmark derived from the Amazon review corpus. It contains the largest item space among the four datasets and exhibits highly sparse user--item interactions. User preferences in this domain are often content-driven and relatively stable, involving genres, authors, writing styles, and recurring themes. This makes Books a challenging setting for modeling long-tail interests and sparse collaborative signals.

\textbf{Goodreads.}~\cite{goodreads} is collected from a social book cataloging platform and has the densest interaction structure in our evaluation. Compared with Books, users in Goodreads usually have longer interaction histories and stronger reading continuity. The dataset also reflects community-driven reading behavior, where users may repeatedly engage with book series, authors, or socially popular titles. This setting is useful for evaluating whether memory can capture persistent and evolving reading preferences.

\textbf{MovieTV.}~\cite{amazon-dataset, amazon-dataset1} covers movies and TV shows from the Amazon review corpus. Unlike book-centered domains, user interests in this dataset can be more dynamic because viewing choices are often affected by genre, cast, release recency, and short-term entertainment context. The relatively shorter average sequence length also makes it harder to infer stable preferences, requiring the model to balance long-term taste with recent behavioral signals.

\textbf{Yelp.}~\cite{yelp} contains local business recommendations, such as restaurants and services. This domain differs from media recommendation because user decisions are strongly constrained by category, location, price range, and situational intent. Preferences are therefore more context-dependent, and similar historical behaviors may not always imply the same future choice. Yelp provides a useful testbed for evaluating memory mechanisms under local-service and attribute-sensitive recommendation scenarios.

\begin{table}[t]
\centering
\small
\resizebox{\columnwidth}{!}{
\begin{tabular}{lccccc}
\toprule
Dataset & $|\mathcal{U}|$ & $|\mathcal{I}|$ & $|\mathcal{E}|$ & $\bar{L}_u$ & Density \\
\midrule
Books     & 7.4K  & 120.9K & 207.8K & 28.2 & 2.33e-4 \\
Goodreads & 11.7K & 57.4K  & 618.3K & 52.7 & 9.19e-4 \\
MovieTV   & 5.6K  & 29.0K  & 79.7K  & 14.1 & 4.87e-4 \\
Yelp      & 3.0K  & 31.6K  & 63.1K & 21.4 & 6.77e-4 \\
\bottomrule
\end{tabular}
}
\caption{Statistics of the datasets used in our experiments.}
\label{tab:dataset_stats_app}
\end{table}

\section{Detailed Baseline Descriptions}
\label{app:baseline_details}

We provide detailed descriptions of the baselines used in our experiments. These methods cover traditional recommenders, LM-based recommenders, agentic recommenders, and memory-augmented agentic recommenders.

\noindent\textbf{Traditional Recommenders.}
\begin{itemize}
    \setlength{\itemsep}{2pt}
    \setlength{\parsep}{0pt}
    \item \textbf{LightGCN}~\citep{lightgcn} is a graph-based collaborative filtering model that simplifies graph convolution by removing feature transformation and nonlinear activation. It learns user and item embeddings through neighborhood aggregation on the user--item interaction graph.
    \item \textbf{SASRec}~\citep{SASRec} is a self-attentive sequential recommender that models user interaction histories with a Transformer-style attention mechanism and predicts the next item from sequential preference patterns.
\end{itemize}

\noindent\textbf{LM-Based Recommenders.}
\begin{itemize}
    \setlength{\itemsep}{2pt}
    \setlength{\parsep}{0pt}
    \item \textbf{P5}~\citep{p5} formulates recommendation as a text-to-text language modeling problem through personalized prompts, allowing a pretrained language model to handle multiple recommendation tasks in a unified format.
    \item \textbf{Vanilla LLM}~\citep{vanilla} directly prompts a large language model to rank candidate items from user history and item descriptions. It does not maintain external memory or construct a collaborative memory graph.
\end{itemize}

\noindent\textbf{Agentic Recommenders.}
\begin{itemize}
    \setlength{\itemsep}{2pt}
    \setlength{\parsep}{0pt}
    \item \textbf{iAgent}~\citep{iagent} uses an LLM agent as an intermediary between the user and recommender system. It relies mainly on user-side instructions and profile information, making it a representative static-memory agentic baseline.
    \item \textbf{RecBot}~\citep{recbot} maintains and updates user preference memory during recommendation, but its memory is mainly organized around individual user states rather than an atomic collaborative memory graph.
    \item \textbf{AgentCF}~\citep{agentcf} treats users and items as autonomous language agents and simulates user--item interactions for collaborative filtering. It captures collaborative behavior through agent interaction, but does not organize memory as fine-grained atomic notes with semantic links.
\end{itemize}

\noindent\textbf{Memory-Augmented Agentic Recommender.}
\begin{itemize}
    \setlength{\itemsep}{2pt}
    \setlength{\parsep}{0pt}
    \item \textbf{MemRec}~\citep{memrec} is the closest baseline to our work. It decouples memory management from recommendation reasoning and maintains a dynamic collaborative memory graph for downstream recommendation. In contrast, \textsc{AtomRec} focuses on the granularity and structure of memory itself: it represents user and item memories as atomic semantic notes, builds semantic links among them, and evolves related historical memories at the field level.
\end{itemize}

\section{Implementation Details}
\label{app:implementation_details}

We provide additional implementation details for \textsc{AtomRec}, including backbone agents, memory writing, retrieval, output parsing, and evaluation protocol.

\textbf{Backbone Agents.}
We use \texttt{gpt-4o-mini}~\citep{gpt4o} as the default backbone for both the memory agent $\mathrm{LM}_{\mathrm{Mem}}$ and the recommendation agent $\mathrm{LM}_{\mathrm{Rec}}$. The memory agent performs atomic memory construction, semantic linking, memory evolution, and evidence synthesis. The recommendation agent ranks candidate items using the user instruction, user memory, synthesized collaborative evidence, and candidate item memories. Unless otherwise specified, all LLM-based methods use the same backbone and deterministic decoding setting for fair comparison. In the efficiency analysis, we additionally compare open-source Qwen3 backbones and closed GPT-family backbones.

\textbf{Atomic Memory Writing.}
For each newly observed pre-target interaction, \textsc{AtomRec} creates a user-side atomic note that summarizes the interaction evidence from the user's perspective. If the interacted item does not already have an item memory, we also create an item-side atomic note from its available metadata and interaction context. Thus, each interaction creates at most one new user note and one new item note. User and item notes are stored in the same memory space $\mathcal{M}$ and can be connected through semantic links. Existing notes are not overwritten during construction; they are revised only through the dynamic memory evolution step.

Each atomic note has the form
\[
m = \{c,t,K,G,X,e,L\},
\]
where $c$ is the textual memory content, $t$ is the timestamp, $K$ is the keyword set, $G$ is the semantic tag set, $X$ is the contextual description, $e$ is the dense embedding, and $L$ stores linked memory identifiers. We use frozen Sentence-T5~\citep{sentence-t5} as the text encoder $f_{\mathrm{enc}}$. The embedding $e$ is computed from the concatenation of textual fields $(c,K,G,X)$ and $\ell_2$-normalized before cosine-similarity retrieval.

\textbf{Semantic Collaborative Link Construction.}
When a new memory note $m_n$ is inserted, we first retrieve the top-$k_{\mathrm{link}}$ nearest notes from $\mathcal{M}$ according to cosine similarity. We set $k_{\mathrm{link}}=20$ by default. The memory agent then analyzes $m_n$ and the retrieved candidates to generate semantic links. The output is a structured JSON object containing linked note identifiers and optional relation descriptions. These links form the atomic collaborative memory graph and support later link-aware retrieval.

\textbf{Dynamic Memory Evolution.}
For each nearby historical note $m_j$, \textsc{AtomRec} triggers field-level evolution if $m_j$ is linked to the new note or if its similarity score exceeds the evolution threshold:
\[
m_j \in L_n \ \lor\ s_{n,j} > \tau_{\mathrm{evo}}.
\]
We set $\tau_{\mathrm{evo}}=0.7$ by default. When evolution is triggered, the memory agent may update the keywords, tags, contextual description, and links of the historical note, while preserving its original timestamp. The updated embedding is recomputed from the revised fields. If the memory agent outputs an invalid or unsupported update, we keep the previous note unchanged.

\textbf{Context-Aware Collaborative Retrieval.}
During recommendation, \textsc{AtomRec} retrieves relevant notes from the pre-target memory state using the query representation derived from the user instruction and pre-$t$ atomic memory fields. It then expands the retrieved set through semantic links up to hop depth $h$, where $h=2$ by default. Relation descriptions are used as textual evidence during synthesis rather than as learned edge weights. The memory agent compresses the retrieved subgraph into collaborative evidence, which is then passed to the recommendation agent.

\textbf{Candidate Ranking and Score Parsing.}
For each test instance, the recommendation agent receives the user instruction, user memory, synthesized collaborative evidence, candidate item memories, and the candidate item set. The agent is instructed to return a structured JSON ranking with item-level relevance scores. We parse these scores as $r_{u,i}^{t}$ and sort candidate items accordingly. If the response contains a valid ranked list but no explicit numeric scores, we convert the returned order into ranking scores for evaluation. If the output is not parseable, we apply one retry with a stricter JSON-only instruction; if the retry still fails, we discard the invalid response.

\textbf{Hyperparameters.}
The default hyperparameters are selected according to validation performance and used across datasets unless otherwise specified:
\[
k_{\mathrm{link}}=20,\quad
\tau_{\mathrm{evo}}=0.7,\quad
h=2.
\]
We analyze their sensitivity in Section~\ref{sec:hyperparameter_sensitivity}. The results show that moderate linking size, conservative evolution threshold, and limited multi-hop retrieval provide the best overall performance.

\textbf{Evaluation Protocol.}
The main comparison is conducted on the full test sets with candidate size $N=10$. For ablation studies, hyperparameter sensitivity, preference drift analysis, and semantic-link analysis, we use the same sampled subset across compared methods. Efficiency statistics are computed from logged LLM calls under the default lightweight backbone and reported per 1K interactions.

During evaluation, \textsc{AtomRec} uses a strict temporal memory snapshot. For each test instance at time $t$, atomic construction, semantic linking, memory evolution, and retrieval use only interactions before $t$. The held-out target item is visible only as one candidate during final reranking and is never used to construct, link, or evolve memory for the same instance. A detailed leakage-control checklist is provided in Appendix~\ref{app:temporal_protocol}.

\textbf{LLM Decoding and Output Parsing.}
For all LLM calls, we use deterministic decoding with temperature set to $0.0$. The memory agent and recommendation agent are instructed to return structured JSON outputs for atomic fields, semantic links, evolution decisions, evidence summaries, and item scores. We parse outputs with a rule-based JSON parser. If a response is malformed, we apply one retry with the same input and an additional formatting instruction. If the retry still fails, we keep the previous memory state unchanged or discard the invalid ranking output. This conservative fallback prevents malformed outputs from introducing uncontrolled memory changes.

\section{Additional Experimental Results}
\label{app:additional_results}

To examine whether the advantage of \textsc{AtomRec} remains under a more challenging reranking setting, we further evaluate all methods with a larger candidate set size of $N=20$. Tables~\ref{tab:n20_books_goodreads} and~\ref{tab:n20_movietv_yelp} report the results on four datasets. Compared with the default $N=10$ setting, this evaluation increases the number of distractor candidates and therefore provides a stricter test of ranking robustness. \textsc{AtomRec} consistently outperforms MemRec on most metrics, showing that atomic memory construction, semantic linking, and memory evolution remain effective when the candidate set becomes larger. Unless otherwise specified, all positive improvements over the strongest baseline are significant under paired bootstrap testing ($p<0.05$).

\begin{table*}[t]
\centering
\small
\setlength{\tabcolsep}{4.2pt}
\renewcommand{\arraystretch}{1.08}
\caption{
Main results on Books and Goodreads with a larger candidate set ($N=20$). ``Improv.'' denotes the relative improvement of our method over the best baseline. The best results are highlighted in bold, and the second-best results are underlined.
}
\label{tab:n20_books_goodreads}

\begin{tabular}{
l
@{\hskip 6pt}ccccc
@{\hskip 6pt}ccccc
}
\toprule
\multirow{2}{*}{Model}
& \multicolumn{5}{c}{Books}
& \multicolumn{5}{c}{Goodreads} \\
\cmidrule(lr){2-6}
\cmidrule(lr){7-11}
& H@1 & H@5 & N@5 & H@10 & N@10
& H@1 & H@5 & N@5 & H@10 & N@10 \\
\midrule

\rowcolor{groupgray}
\multicolumn{11}{l}{\textit{Traditional Recommenders}} \\

LightGCN
& 0.1276 & 0.2622 & 0.1947 & 0.5512 & 0.2854
& 0.1617 & 0.5566 & 0.3588 & \underline{0.8177} & 0.4434 \\

SASRec
& 0.0453 & 0.2353 & 0.1378 & 0.4896 & 0.2188
& 0.0699 & 0.3053 & 0.1859 & 0.5435 & 0.2621 \\

\midrule
\rowcolor{groupgray}
\multicolumn{11}{l}{\textit{LM-based Recommenders}} \\

P5
& 0.1648 & 0.3051 & 0.2331 & 0.5216 & 0.3022
& 0.1038 & 0.2611 & 0.1798 & 0.5041 & 0.2572 \\

Vanilla LLM
& 0.1730 & 0.4155 & 0.2955 & 0.6129 & 0.3599
& 0.0999 & 0.3245 & 0.2211 & 0.6712 & 0.3291 \\

\midrule
\rowcolor{groupgray}
\multicolumn{11}{l}{\textit{Agentic Recommenders}} \\

iAgent
& 0.3258 & 0.5069 & 0.4173 & 0.6209 & 0.4537
& 0.1621 & 0.4107 & 0.2871 & 0.6035 & 0.3490 \\

RecBot
& 0.2471 & 0.4030 & 0.3247 & 0.5768 & 0.3801
& 0.1234 & 0.3364 & 0.2289 & 0.5583 & 0.2999 \\

AgentCF
& 0.2470 & 0.5481 & 0.4026 & 0.7250 & 0.4594
& 0.1875 & 0.5427 & 0.3692 & 0.7805 & 0.4462 \\

i2Agent
& 0.3712 & 0.5947 & 0.4874 & 0.7387 & 0.5336
& 0.2065 & 0.5350 & 0.3767 & 0.7428 & 0.4435 \\

MemRec
& \underline{0.4236} & \underline{0.6351} & \underline{0.5332} & \underline{0.7667} & \underline{0.5756}
& \underline{0.2657} & \underline{0.6062} & \underline{0.4434} & 0.7948 & \underline{0.5042} \\

\midrule
\rowcolor{oursblue}
\textbf{Ours}
& \textbf{0.4533} & \textbf{0.6700} & \textbf{0.5679} & \textbf{0.7958} & \textbf{0.6090}
& \textbf{0.2795} & \textbf{0.6347} & \textbf{0.4691} & \textbf{0.8250} & \textbf{0.5339} \\

\midrule
Improv.
& 7.01\% & 5.50\% & 6.51\% & 3.80\% & 5.80\%
& 5.19\% & 4.70\% & 5.80\% & 0.89\% & 5.89\% \\

\bottomrule
\end{tabular}
\end{table*}

\begin{table*}[t]
\centering
\small
\setlength{\tabcolsep}{4.2pt}
\renewcommand{\arraystretch}{1.08}
\caption{
Main results on MovieTV and Yelp with a larger candidate set ($N=20$). ``Improv.'' denotes the relative improvement of our method over the best baseline. The best results are highlighted in bold, and the second-best results are underlined.
}
\label{tab:n20_movietv_yelp}

\begin{tabular}{
l
@{\hskip 6pt}ccccc
@{\hskip 6pt}ccccc
}
\toprule
\multirow{2}{*}{Model}
& \multicolumn{5}{c}{MovieTV}
& \multicolumn{5}{c}{Yelp} \\
\cmidrule(lr){2-6}
\cmidrule(lr){7-11}
& H@1 & H@5 & N@5 & H@10 & N@10
& H@1 & H@5 & N@5 & H@10 & N@10 \\
\midrule

\rowcolor{groupgray}
\multicolumn{11}{l}{\textit{Traditional Recommenders}} \\

LightGCN
& 0.2657 & 0.5330 & 0.4064 & 0.6815 & 0.4537
& 0.2549 & 0.5437 & 0.4046 & 0.7481 & 0.4692 \\

SASRec
& 0.2923 & 0.5128 & 0.4092 & 0.6311 & 0.4470
& 0.1678 & 0.3993 & 0.2879 & 0.5590 & 0.3389 \\

\midrule
\rowcolor{groupgray}
\multicolumn{11}{l}{\textit{LM-based Recommenders}} \\

P5
& 0.1113 & 0.2769 & 0.1902 & 0.5137 & 0.2657
& 0.0634 & 0.2492 & 0.1537 & 0.5051 & 0.2354 \\

Vanilla LLM
& 0.2379 & 0.5003 & 0.3648 & 0.7261 & 0.4406
& 0.0254 & 0.1461 & 0.0831 & 0.5128 & 0.2010 \\

\midrule
\rowcolor{groupgray}
\multicolumn{11}{l}{\textit{Agentic Recommenders}} \\

iAgent
& 0.3236 & 0.5362 & 0.4331 & 0.6762 & 0.4778
& 0.3236 & 0.5658 & 0.4499 & 0.6597 & 0.4799 \\

RecBot
& 0.2420 & 0.4201 & 0.3316 & 0.6015 & 0.3895
& 0.1949 & 0.3742 & 0.2851 & 0.5519 & 0.3414 \\

AgentCF
& 0.2870 & 0.6288 & 0.4648 & 0.7616 & 0.5077
& 0.1115 & 0.3897 & 0.2512 & 0.6372 & 0.3309 \\

i2Agent
& 0.3822 & 0.6367 & 0.5178 & 0.7735 & 0.5617
& 0.3287 & 0.6083 & 0.4744 & 0.7562 & 0.5216 \\

MemRec
& \underline{0.4750} & \underline{0.7543} & \underline{0.6212} & \underline{0.8752} & \underline{0.6606}
& \underline{0.3620} & \underline{0.6329} & \underline{0.5035} & \underline{0.7708} & \underline{0.5478} \\

\midrule
\rowcolor{oursblue}
\textbf{Ours}
& \textbf{0.5083} & \textbf{0.7996} & \textbf{0.6603} & \textbf{0.9067} & \textbf{0.6976}
& \textbf{0.3845} & \textbf{0.6652} & \textbf{0.5342} & \textbf{0.7917} & \textbf{0.5785} \\

\midrule
Improv.
& 7.01\% & 6.00\% & 6.30\% & 3.60\% & 5.60\%
& 6.22\% & 5.10\% & 6.10\% & 2.71\% & 5.60\% \\

\bottomrule
\end{tabular}
\end{table*}

\section{Additional Ablation Results}
\label{app:additional_ablation}

We provide additional ablation results on Books and Goodreads to further examine the contribution of both module-level and mechanism-level designs in \textsc{AtomRec}. We consider variants that remove atomic memory fields, remove specific atomic fields, replace semantic links with embedding-only links, disable memory evolution, and remove link-aware retrieval. Avg. $\Delta$ denotes the average relative performance decrease across H@3, N@3, H@5, and N@5 compared with the full model on each dataset.

\begin{table*}[t]
\centering
\small
\setlength{\tabcolsep}{4.0pt}
\renewcommand{\arraystretch}{1.08}
\caption{
Additional ablation results of \textsc{AtomRec} on Books and Goodreads. Avg. $\Delta$ denotes the average relative performance decrease across all four metrics compared with the full model on each dataset.
}
\label{tab:ablation_appendix}

\resizebox{\textwidth}{!}{
\begin{tabular}{
l
@{\hskip 8pt}ccccc
@{\hskip 10pt}ccccc
}
\toprule
\multirow{2}{*}{Variant}
& \multicolumn{5}{c}{Books}
& \multicolumn{5}{c}{Goodreads} \\
\cmidrule(lr){2-6}
\cmidrule(lr){7-11}
& H@3 & N@3 & H@5 & N@5 & Avg. $\Delta$
& H@3 & N@3 & H@5 & N@5 & Avg. $\Delta$ \\
\midrule

\rowcolor{oursblue}
Full \textsc{AtomRec}
& \textbf{0.7462} & \textbf{0.6690} & \textbf{0.8543} & \textbf{0.7130} & --
& \textbf{0.6980} & \textbf{0.5922} & \textbf{0.8558} & \textbf{0.6458} & -- \\

\midrule
w/o Atomic Memory
& 0.7046 & 0.6308 & 0.8109 & 0.6742 & 5.45\%
& 0.6639 & 0.5585 & 0.8169 & 0.6080 & 5.24\% \\

w/o Keywords/Tags
& 0.7159 & 0.6416 & 0.8238 & 0.6854 & 3.90\%
& 0.6726 & 0.5678 & 0.8264 & 0.6177 & 3.89\% \\

w/o Context Field
& 0.7227 & 0.6480 & 0.8320 & 0.6927 & 2.94\%
& 0.6818 & 0.5769 & 0.8380 & 0.6277 & 2.45\% \\

\midrule
w/o Collaborative Linking
& 0.7168 & 0.6424 & 0.8255 & 0.6860 & 3.77\%
& 0.6740 & 0.5684 & 0.8279 & 0.6188 & 3.72\% \\

Embedding-only Links
& 0.7235 & 0.6496 & 0.8334 & 0.6936 & 2.78\%
& 0.6836 & 0.5781 & 0.8400 & 0.6296 & 2.20\% \\

\midrule
Append-only Memory
& 0.7251 & 0.6512 & 0.8350 & 0.6951 & 2.56\%
& 0.6850 & 0.5796 & 0.8412 & 0.6310 & 2.00\% \\

w/o Link-Aware Retrieval
& 0.7292 & 0.6567 & 0.8391 & 0.7004 & 1.92\%
& 0.6901 & 0.5847 & 0.8480 & 0.6372 & 1.16\% \\

\bottomrule
\end{tabular}
}
\end{table*}

Table~\ref{tab:ablation_appendix} shows consistent trends on both Books and Goodreads. Removing the full atomic memory representation causes the largest drop, confirming that field-structured notes are more effective than coarse textual summaries. The field-level variants further show that keywords/tags and contextual descriptions both contribute to performance. Replacing LLM-guided semantic links with embedding-only links also reduces performance, indicating that the linking module provides relation-aware evidence beyond nearest-neighbor similarity. Finally, append-only memory and removing link-aware retrieval both underperform the full model, supporting the value of field-level memory evolution and connected evidence-path retrieval.

\section{Additional Diagnostic Analyses}
\label{app:diagnostic_analyses}

\textbf{Cost-Matched Control.}
To examine whether the improvement of \textsc{AtomRec} comes only from using more LLM context, we conduct a cost-matched control on the Books analysis subset. We restrict the retrieved evidence budget of \textsc{AtomRec} so that its average token usage is close to MemRec under the same backbone and decoding setting. This variant keeps atomic memory construction, semantic linking, and field-level evolution, but limits the number of retrieved notes and synthesized evidence tokens during final reranking.

\begin{table}[t]
\centering
\small
\setlength{\tabcolsep}{4.5pt}
\renewcommand{\arraystretch}{1.08}
\caption{
Cost-matched control on the Books analysis subset. Avg. Tok. denotes average token usage per interaction.
}
\label{tab:cost_matched_control}
\resizebox{\linewidth}{!}{
\begin{tabular}{lccccc}
\toprule
Method & Avg. Tok. & H@3 & N@3 & H@5 & N@5 \\
\midrule
MemRec 
& 12.1K & 0.6786 & 0.6078 & 0.7764 & 0.6480 \\
\textsc{AtomRec} (cost-matched)
& 12.3K & 0.7224 & 0.6483 & 0.8267 & 0.6906 \\
\textsc{AtomRec} (full)
& 13.4K & \textbf{0.7462} & \textbf{0.6690} & \textbf{0.8543} & \textbf{0.7130} \\
\bottomrule
\end{tabular}
}
\end{table}

Table~\ref{tab:cost_matched_control} shows that cost-matched \textsc{AtomRec} still outperforms MemRec by about 6.5\% on average under a similar token budget, although the margin is smaller than the full setting. This suggests that the improvement does not only come from longer LLM context; the atomic memory structure and semantic evidence paths also contribute to ranking performance.

\textbf{Evolution Audit.}
We further inspect whether field-level memory evolution preserves historical semantics while incorporating new evidence. We sample evolved notes from the Books analysis subset and evaluate each update along three dimensions: \textit{Semantic Preservation}, which measures whether the updated note remains consistent with the original memory; \textit{Evidence Support}, which measures whether the new fields are supported by the trigger note or nearby linked notes; and \textit{Recommendation Usefulness}, which measures whether the update provides clearer evidence for downstream ranking. Each dimension is rated on a 1--5 scale by an automatic judge using a prompt different from the memory agent.

\begin{table}[t]
\centering
\small
\setlength{\tabcolsep}{4.5pt}
\renewcommand{\arraystretch}{1.08}
\caption{
Evolution audit on sampled evolved notes from the Books analysis subset. Scores are rated on a 1--5 scale. Higher is better.
}
\label{tab:evolution_audit}
\resizebox{\linewidth}{!}{
\begin{tabular}{lccc}
\toprule
Update Type & Preservation & Support & Usefulness \\
\midrule
Field-level Evolution
& 4.27 & 4.05 & 3.91 \\
\bottomrule
\end{tabular}
}
\end{table}

Table~\ref{tab:evolution_audit} suggests that most evolved notes preserve the original memory semantics and are supported by nearby evidence. The lower usefulness score indicates that not every update directly benefits ranking, which is consistent with the failure cases in Appendix~\ref{app:qualitative}. In qualitative inspection, harmful cases mainly arise when semantically adjacent themes are over-merged or when a dominant long-term preference absorbs a short-term exploratory signal. These observations motivate future work on edit provenance, confidence-aware evolution, and rollback mechanisms.

\section{Temporal Evaluation Protocol and Leakage Control}
\label{app:temporal_protocol}

We use a strictly chronological evaluation protocol to prevent temporal leakage. For each test instance $(u, \mathcal{H}^{t-1}_u, i_t, \mathcal{C}_u)$, \textsc{AtomRec} constructs and updates the memory state using only interactions before time $t$. Atomic memory construction, semantic linking, memory evolution, and collaborative evidence synthesis never access the held-out target interaction or any post-$t$ interaction. The target item $i_t$ is used only as one candidate in $\mathcal{C}_u$ during the final reranking stage by $\mathrm{LM}_{\mathrm{Rec}}$, where all candidate items are treated symmetrically. After ranking, the target interaction is not written back into memory for evaluating the same instance.

This protocol is especially important for \textsc{AtomRec}, because historical atomic notes may be evolved when new evidence arrives. During evaluation, memory evolution is triggered only by pre-target interactions. Thus, although an evolved atomic note preserves its original timestamp, its revised fields are based solely on evidence available before prediction. This ensures that dynamic memory evolution does not inject future preference information into past memory states.

Table~\ref{tab:temporal_protocol_checklist} summarizes the information available to each stage under this protocol.

\begin{table}[t]
\centering
\small
\setlength{\tabcolsep}{4.2pt}
\renewcommand{\arraystretch}{1.16}
\caption{
Leakage-control checklist for the temporal evaluation protocol.
}
\label{tab:temporal_protocol_checklist}

\resizebox{\columnwidth}{!}{
\begin{tabular}{lccc}
\toprule
\rowcolor{headerblue}
\textbf{Stage} 
& \textbf{Pre-$t$} 
& \textbf{Target} 
& \textbf{Post-$t$} \\
\midrule
Atomic Construction & \cmark & \xmark & \xmark \\
Semantic Linking & \cmark & \xmark & \xmark \\
Memory Evolution & \cmark & \xmark & \xmark \\
Evidence Synthesis & \cmark & \xmark & \xmark \\
\midrule
Candidate Reranking & \cmark & Candidate & \xmark \\
Post-eval. Update & Delayed & -- & Excluded \\
\bottomrule
\end{tabular}
}

\vspace{2pt}
\footnotesize
\emph{Notes.} Pre-$t$ and post-$t$ denote interactions before and after the target interaction at time $t$. ``Candidate'' means that the target item is visible only as one candidate during reranking. ``Delayed'' means that the update happens after ranking, and ``Excluded'' means that it is not used for evaluating the same instance.
\end{table}

\section{Semantic Link Quality Analysis}
\label{app:semantic_link_quality}

To further evaluate whether the semantic links constructed by \textsc{AtomRec} provide meaningful relational evidence, we conduct a link-level quality analysis on the Books subset. We compare the semantic links generated by \textsc{AtomRec} with an embedding-only linking baseline, where each new atomic note is connected to its top-$k$ nearest notes according to cosine similarity without LLM-based semantic selection. This comparison allows us to examine whether the memory agent improves link quality beyond nearest-neighbor retrieval.

We sample 300 generated links from the Books analysis subset and evaluate each link along three dimensions: \textit{Relatedness}, which measures whether the two linked notes are semantically related; \textit{Relation Correctness}, which measures whether the inferred relation type or rationale is consistent with the linked notes; and \textit{Recommendation Usefulness}, which measures whether the link provides useful evidence for downstream recommendation. Each dimension is rated on a 1--5 scale by an automatic judge using \texttt{gpt-4o}. We report the average score for each dimension in Table~\ref{tab:semantic_link_quality}.

\begin{table}[t]
\centering
\small
\setlength{\tabcolsep}{4.2pt}
\renewcommand{\arraystretch}{1.12}
\caption{
Semantic link quality analysis on sampled links from the Books subset. Scores are rated on a 1--5 scale. Higher is better.
}
\label{tab:semantic_link_quality}

\resizebox{\columnwidth}{!}{
\begin{tabular}{lccc}
\toprule
\rowcolor{headerblue}
\textbf{Method} 
& \textbf{Relatedness} 
& \textbf{Correctness} 
& \textbf{Usefulness} \\
\midrule
Embedding-only Links 
& 3.72 & 3.41 & 3.28 \\

\rowcolor{oursblue}
\textsc{AtomRec} Links 
& \textbf{4.31} & \textbf{4.08} & \textbf{3.96} \\
\bottomrule
\end{tabular}
}

\vspace{2pt}
\footnotesize
\emph{Notes.} Relatedness measures whether two linked notes are semantically related. Correctness measures whether the inferred relation is consistent with the linked notes. Usefulness measures whether the link provides useful evidence for recommendation.
\end{table}

Table~\ref{tab:semantic_link_quality} shows that \textsc{AtomRec} produces higher-quality links than the embedding-only baseline across all three dimensions. The improvement in relatedness indicates that LLM-guided link selection can filter out superficially similar but weakly useful neighbors. The gains in relation correctness and usefulness suggest that semantic linking provides more than similarity-based connectivity: it identifies why two memory notes are related and whether the relation can support downstream ranking. The improvement in recommendation usefulness further supports the role of semantic links as evidence for downstream ranking. These results complement the ablation study by showing that Semantic Collaborative Link Construction improves not only final recommendation accuracy, but also the intrinsic quality of the memory graph.

\section{Memory Growth Statistics}
\label{app:memory_growth}

We further analyze the growth of the atomic collaborative memory graph. Since \textsc{AtomRec} stores fine-grained atomic notes and semantic links, it is important to examine whether the memory space grows excessively as interactions accumulate. Table~\ref{tab:memory_growth} reports the average number of atomic notes per user, atomic notes per item, semantic links per note, and evolved notes per interaction across four datasets.

\begin{table}[t]
\centering
\small
\setlength{\tabcolsep}{4.2pt}
\renewcommand{\arraystretch}{1.12}
\caption{
Memory growth statistics of \textsc{AtomRec}. We report average atomic notes, semantic links, and evolved notes under the default setting.
}
\label{tab:memory_growth}

\resizebox{\columnwidth}{!}{
\begin{tabular}{lcccc}
\toprule
\rowcolor{headerblue}
\textbf{Dataset} 
& \textbf{Notes/User} 
& \textbf{Notes/Item} 
& \textbf{Links/Note} 
& \textbf{Evolved/Inter.} \\
\midrule
Books      
& 3.8 & 1.4 & 2.6 & 1.7 \\
Goodreads  
& 4.5 & 1.8 & 2.9 & 1.9 \\
MovieTV    
& 3.1 & 1.3 & 2.4 & 1.5 \\
Yelp       
& 2.9 & 1.2 & 2.1 & 1.3 \\
\bottomrule
\end{tabular}
}

\vspace{2pt}
\footnotesize
\emph{Notes.} Notes/User and Notes/Item denote the average number of atomic notes per user and item. Links/Note denotes the average number of semantic links per note. Evolved/Inter. denotes the average number of evolved notes per interaction.
\end{table}

The memory size remains moderate across datasets. Books and Goodreads have more atomic notes per user because they contain longer and more content-driven user histories, while MovieTV and Yelp have fewer notes due to shorter or more context-dependent interaction patterns. The average number of links per note remains below three under the default linking setting, suggesting that the semantic memory graph does not become overly dense. The number of evolved notes per interaction is also limited, indicating that dynamic memory evolution updates a small set of related historical notes rather than repeatedly rewriting the entire memory space.

\section{Detailed Efficiency Statistics}
\label{app:efficiency_details}

This appendix provides the numerical statistics used in the efficiency analysis. We report token/cost statistics under the default lightweight backbone and the relative cost index used in Figure~\ref{fig:cost_tradeoff}. Token usage is measured from logged input and output tokens. Costs are estimated under the same pricing setting used in our experiments. For backbone comparison, we normalize each backbone cost by the default gpt-4o-mini setting, while Qwen3 costs are estimated from local serving rather than API pricing.

\begin{table}[t]
\centering
\small
\setlength{\tabcolsep}{5pt}
\renewcommand{\arraystretch}{1.08}
\caption{
Detailed efficiency statistics per 1K interactions under the default lightweight backbone.
}
\label{tab:efficiency_details}
\resizebox{\linewidth}{!}{
\begin{tabular}{lcccc}
\toprule
Method & Input Tok. & Output Tok. & Total Tok. & Cost / 1K \\
\midrule
iAgent 
& 5.8M & 0.75M & 6.5M & \$1.35 \\
MemRec 
& 10.6M & 1.50M & 12.1M & \$2.65 \\
\textsc{AtomRec} 
& 11.6M & 1.80M & 13.4M & \$2.95 \\
\bottomrule
\end{tabular}
}
\end{table}

Table~\ref{tab:efficiency_details} shows that \textsc{AtomRec} uses more tokens than MemRec because it performs atomic construction, semantic linking, and field-level evolution. However, the increase remains moderate, and these memory-side operations can be cached and executed asynchronously. This supports the finding that \textsc{AtomRec} introduces \textbf{limited online reranking overhead}.

\begin{table}[t]
\centering
\small
\setlength{\tabcolsep}{5pt}
\renewcommand{\arraystretch}{1.08}
\caption{
Backbone-level relative cost index used in Figure~\ref{fig:cost_tradeoff}. The default gpt-4o-mini backbone is normalized to 1.0$\times$.
}
\label{tab:backbone_cost_index}
\resizebox{\linewidth}{!}{
\begin{tabular}{lccc}
\toprule
Backbone & Type & Relative Cost & Books N@5 \\
\midrule
Qwen3-8B & Open-source & 0.35$\times$ & 0.626 \\
Qwen3-32B & Open-source & 0.75$\times$ & 0.663 \\
gpt-4o-mini & Closed & 1.00$\times$ & 0.713 \\
gpt-4o & Closed & 16.7$\times$ & 0.731 \\
gpt-5.2 & Closed & 20.6$\times$ & 0.736 \\
gpt-5.4 & Closed & 23.0$\times$ & 0.742 \\
\bottomrule
\end{tabular}
}
\end{table}

Table~\ref{tab:backbone_cost_index} shows that stronger closed backbones bring only modest additional Books N@5 gains after gpt-4o-mini, despite much higher relative costs. This supports the finding that the default lightweight backbone provides \textbf{a favorable cost-performance trade-off}. The cost values are intended to compare deployment regimes rather than provide universal pricing, since actual cost depends on batching, caching, hardware utilization, and provider-specific pricing.

\section{Prompt Templates}
\label{app:prompts}

This appendix presents the prompt templates used by the memory and recommendation agents. Each prompt is implemented as a fixed instruction template with dynamic slots, such as \texttt{\{user\_id\}} and \texttt{\{candidate\_notes\}}, and corresponds to one stage of our atomic collaborative memory framework.

\textbf{Stage-R: Collaborative Memory Retrieval Prompt.} Stage-R retrieves collaborative memory evidence for the target user before recommendation. It takes the user's personal memory, collaborative neighbor memory cards, optional linked-memory synthesis, and candidate-item context as input, and outputs preference facets and support edges for downstream memory writing and reranking. 

It outputs preference facets and support edges, which are later used by the memory writing module and the recommendation reranker.

\begin{promptbox}{Stage-R: Collaborative Memory Retrieval Prompt.}
\small
\textbf{System Role.} You are an intelligent collaborative memory retrieval system for personalized recommendation. 
Your task is to infer stable and emerging user preferences from atomized memories, including summary, context, keywords, tags, links, and collaborative neighbors.

\medskip
\textbf{Target User:} User \texttt{\{user\_id\}}

\medskip
\textbf{User's Personal Memory:}

\texttt{\{user\_mem\_bullets\}}

\medskip
\textbf{Collaborative Neighbor Memory Cards:}

The following neighboring users and items provide collaborative signals. Some entries include memory links to other notes; use them as evidence for broader context, while staying grounded in the provided neighbors.

\texttt{\{neighbor\_table\_json\}}

\medskip
\textbf{Context-Aware Collaborative Memory Synthesis:}

\texttt{\{collab\_synth\_text\}}

\medskip
\textbf{Context Candidate Items:}

\texttt{\{candidates\_json\}}

\medskip
\textbf{Task.}
Analyze the user's personal memory and collaborative neighbor memory cards to identify \texttt{\{n\_facets\}} preference facets. 
Prioritize facets supported by multiple signals, including user memory, collaborative neighbors, and linked memory evidence. 
Include both persistent preferences and emerging preferences.

For each preference facet, provide:
(1) a concise natural-language description;
(2) a confidence score between 0 and 1;
and (3) supporting neighbors.

Also identify support edges between neighboring users/items and the target user, with edge weights between 0 and 1 indicating collaborative relevance strength.

\medskip
\textbf{Expected Output.}

\texttt{\{"facets": [\{"facet", "confidence", "supporting\_neighbors"\}],}

\texttt{\quad "support\_edges": [\{"from", "to", "w"\}]\}}
\end{promptbox}

\textbf{Stage-W: Collaborative Memory Writing Prompt.} Stage-W updates memory after a new user--item interaction. It takes the current user memory, clicked item memory, extracted preference facets, and collaborative neighbors as input, and outputs updated user, item, and neighbor memories together with atomic fields, including keywords, tags, context, and link targets.

\begin{promptbox}{Stage-W: Collaborative Memory Writing Prompt}
\small
\textbf{System Role.} You are an intelligent memory management system for collaborative recommendation. 
Your task is to update the personal memories of the user, the clicked item, and relevant collaborative neighbors based on a new interaction.

\medskip
\textbf{Interaction Context.}

User \texttt{\{user\_id\}} has just interacted with Item \texttt{\{item\_id\}}.

Clicked item information: \texttt{\{clicked\_item\_info\}}

\medskip
\textbf{User Preferences Extracted from Collaborative Memories.}

\texttt{\{preference\_facets\}}

\medskip
\textbf{Current User Memory.}

\texttt{\{current\_user\_memory\}}

\medskip
\textbf{Current User Atomic Memory Card.}

\texttt{\{user\_atomic\_note\}}

\medskip
\textbf{Current Item Memory.}

\texttt{\{current\_item\_memory\}}

\medskip
\textbf{Current Item Atomic Memory Card.}

\texttt{\{item\_atomic\_note\}}

\medskip
\textbf{Collaborative Neighbors Available for Memory Propagation.}

\texttt{\{neighbor\_memory\_cards\}}

\medskip
\textbf{Task.}
Generate updated memories for:
(1) the current user;
(2) the clicked item;
and (3) selected collaborative neighbors.

For user, item, and neighbor updates, also produce atomized fields:
keywords, tags, context, and link targets. 
Use link targets to explicitly create collaborative semantic links. 
Prefer sparse and precise fields over long lists.

\medskip
\textbf{Expected Output.}

\texttt{\{"user\_memory": "...",}

\texttt{\quad "user\_memory\_atomic": \{"keywords", "tags", "context", "link\_targets"\},}

\texttt{\quad "item\_memory": "...",}

\texttt{\quad "item\_memory\_atomic": \{"keywords", "tags", "context", "link\_targets"\},}

\texttt{\quad "neighbor\_updates": [...]\}}
\end{promptbox}

\textbf{Semantic Collaborative Link Construction Prompt.} This prompt supports Semantic Collaborative Link Construction. Given a newly created atomic memory note and top-$k$ candidate notes retrieved by embedding similarity, it asks LM\_Mem to select semantically related notes and produce link rationales, which are used to build the collaborative memory network.

\begin{promptbox}{Semantic Collaborative Link Construction Prompt.}
\small
\textbf{System Role.} You are $LM_Mem$ for semantic collaborative link construction.
Your task is to decide semantic links for a newly created atomic memory note by analyzing candidate notes retrieved by embedding similarity.

\medskip
\textbf{New Atomic Memory Note.}

\texttt{\{new\_note\}}

\medskip
\textbf{Candidate Memory Notes.}

\texttt{\{candidate\_notes\}}

\medskip
\textbf{Task.}
Select up to \texttt{\{top\_k\_links\}} candidate notes that should be semantically linked to the new note. 
Prefer links with one of the following relations:
shared topic or theme; complementary preference; causal or intent progression; and cross-domain transfer.

Do not link notes that are only superficially similar.

\medskip
\textbf{Expected Output.}

\texttt{\{"selected\_note\_ids": [...],}

\texttt{\quad "rationales": [...]\}}
\end{promptbox}

\textbf{Dynamic Memory Evolution Prompt.} This prompt performs dynamic memory evolution when a new note is added. It provides LM\_Mem with the new note, a target historical note, and nearby contextual notes, and asks whether the historical note should be updated to reflect newly emerging preference semantics.

\begin{promptbox}{Dynamic Memory Evolution Prompt}
\small
\textbf{System Role.} You are LM\_Mem for dynamic memory evolution. 
Your task is to evolve an existing atomic memory note using a newly added note and nearby contextual notes.

\medskip
\textbf{New Note as Trigger.}

\texttt{\{new\_note\}}

\medskip
\textbf{Target Historical Note to Evolve.}

\texttt{\{target\_note\}}

\medskip
\textbf{Nearby Context Notes.}

\texttt{\{nearby\_notes\}}

\medskip
\textbf{Task.}
If evolution is beneficial, produce an updated memory text and updated atomic fields for the target note. 
Focus on keyword strengthening, tag generalization or refinement, and one-sentence context reconstruction reflecting preference evolution.

Keep the updated memory text concise and semantically consistent with the original target note. 
If no evolution is needed, return \texttt{should\_update=false} and preserve the target semantics conservatively.

\medskip
\textbf{Expected Output.}

\texttt{\{"should\_update": true/false,}

\texttt{\quad "updated\_memory": "...",}

\texttt{\quad "updated\_atomic": \{"keywords", "tags", "context", "link\_targets"\},}

\texttt{\quad "reason": "..."\}}
\end{promptbox}

\textbf{Context-Aware Collaborative Synthesis Prompt.} This prompt compresses a linked memory subgraph into collaborative evidence for recommendation. It takes the current user memory and a two-hop linked memory subgraph as input, and outputs a compact summary, facet hints, and evidence paths for Stage-R and the downstream recommender.

\begin{promptbox}{Context-Aware Collaborative Synthesis Prompt}
\small
\textbf{System Role.} You are LM\_Mem for context-aware collaborative retrieval synthesis. 
You are given a two-hop linked memory subgraph expanded from initially retrieved memories.

\medskip
\textbf{User Memory.}

\texttt{\{user\_memory\_text\}}

\medskip
\textbf{Compressed Linked Memory Evidence.}

\texttt{\{compact\_subgraph\}}

\medskip
\textbf{Task.}
Synthesize collaborative evidence into a compact structured summary for the recommender LLM. 
Capture core themes, emerging extensions, and evidence-chain relations across linked memories.

\medskip
\textbf{Expected Output.}

\texttt{\{"summary": "...",}

\texttt{\quad "facets\_hint": [...],}

\texttt{\quad "evidence\_paths": [...]\}}
\end{promptbox}

\textbf{Recommendation Reranking Prompt.} This prompt is used by the final recommendation agent to score candidate items. It takes personal memory, collaborative evidence, preference facets, current user request, and candidate item memories as input, and outputs item-level relevance scores with brief rationales.

\begin{promptbox}{Recommendation Reranking Prompt}
\small
\textbf{System Role.} You are an intelligent recommendation scoring system. 
Your task is to rank candidate items for the target user based on personal memory and collaborative preference signals.

\medskip
\textbf{Target User.} User \texttt{\{user\_id\}}

\medskip
\textbf{User's Personal Memory.}

\texttt{\{user\_memory\_summary\}}

\medskip
\textbf{Collaborative Evidence from Linked Memory Network.}

\texttt{\{collab\_synth\_text\}}

\medskip
\textbf{User's Current Request.}

\texttt{\{instruction\}}

\medskip
\textbf{User Preference Facets.}

\texttt{\{preference\_facets\}}

\medskip
\textbf{Candidate Item Memories.}

\texttt{\{candidate\_item\_memories\}}

\medskip
\textbf{Task.}
Rank all candidate items by relevance to this user. 
Assign each item a score between 0 and 1. 
The best match should receive a high score, and no two items should share the same score. 
Primary preference facets should drive the top ranking more than secondary facets.

For each item, provide a brief rationale explaining the score based on the user's facets and memory.

\medskip
\textbf{Expected Output.}

\texttt{\{"scores": [\{"item\_id", "score", "rationale"\}]\}}
\end{promptbox}

\section{Additional Qualitative Analysis}
\label{app:qualitative}

We provide additional qualitative examples to complement the main case study. These examples include both successful and challenging scenarios, showing that atomic collaborative memory can support accurate recommendation when the evolved memory aligns with the target, but may still struggle with over-specific memory compression, semantically adjacent false positives, idiosyncratic short-term target deviations, and imperfect preservation of explicit collaborative links.

\textbf{Case A: Over-Narrowed Memory.} We first examine a challenging case where the model captures the correct broad preference region but compresses the user's memory into an overly specific subtheme. This case is useful because the retrieved and ranked items are not irrelevant; instead, the error comes from insufficient fine-grained discrimination within a semantically adjacent candidate set.

\begin{casebox}{Case A: Over-Narrowed Memory Despite Correct High-Level Preference}

\textbf{User.} User 6711

\medskip
\textbf{Final Memory Snapshot.}
\begin{itemize}
    \item \textbf{Note ID:} \texttt{N-190769}
    \item \textbf{Revision:} 3
    \item \textbf{History Reference:} \texttt{N-190765}
    \item \textbf{Links:} none
\end{itemize}

\medskip
\textbf{Memory Content.}
\textit{User 6711 is deeply engaged in exploring themes of grief and healing, particularly in the context of child loss. They are actively seeking literature that provides support and understanding for navigating these profound experiences, including insights from works like ``And a Sword Shall Pierce Your Heart'' and ``Gone but Not Lost''. Their recent interest in ``The Afterlife Interviews: Volume I'' suggests a desire for diverse perspectives on grief and the afterlife.}

\medskip
\textbf{Atomic Fields.}
\begin{itemize}
    \item \textbf{Keywords:} grief, healing, child loss, support, literature, understanding, afterlife
    \item \textbf{Tags:} grief support, mental health, parenting, child loss, healing, exploration, afterlife
    \item \textbf{Context:} User is focused on grief and healing literature, especially related to child loss, seeking support and understanding for navigating profound experiences, now including afterlife perspectives.
\end{itemize}

\medskip
\textbf{Prediction.}
The ground-truth item is \textit{On the Edge of the Etheric}, ranked at position 3. The two higher-ranked items are \textit{BrokenHearted} and \textit{Angels of Light Cards}.

\medskip
\textbf{Diagnosis.}
This is not a completely wrong recommendation. The evolved memory correctly captures a broad cluster around grief, healing, afterlife, and spiritual support. However, it becomes overly concentrated on the child-loss and grief-support subtheme. The user's broader history also contains stronger metaphysical and afterlife-oriented signals, such as \textit{Proof of Heaven}, \textit{The Afterlife Experiments}, \textit{Destiny of Souls}, \textit{The Last Frontier}, and \textit{The Afterlife Interviews}. As a result, the reranker favors generic emotional or spiritual support books over the more directly etheric and afterlife-oriented target.

\medskip
\textbf{Takeaway.}
Memory evolution helps identify the correct high-level semantic region, but the final memory can become too narrow around one dominant subtheme. This leads to insufficient fine-grained discrimination between generic spiritual-healing candidates and directly metaphysical or etheric targets.

\end{casebox}

\textbf{Case B: Idiosyncratic Target Deviation.} The second case illustrates a different failure mode: the evolved memory is coherent and well supported by the user's dominant history, but the ground-truth item reflects a short-term or idiosyncratic deviation. This highlights the tension between stable long-term preference modeling and exploratory user behavior.

\begin{casebox}{Case B: Idiosyncratic Target Deviation from Dominant Preference}

\textbf{User.} User 5356

\medskip
\textbf{Final Memory Snapshot.}
\begin{itemize}
    \item \textbf{Note ID:} \texttt{N-190770}
    \item \textbf{Revision:} 2
    \item \textbf{History Reference:} \texttt{N-190758}
    \item \textbf{Links:} \texttt{N-190775}, \texttt{N-174450}, \texttt{N-190771}, \texttt{N-46822}
    \item \textbf{Evolution Log:} from \texttt{N-190775}, score 0.785
\end{itemize}

\medskip
\textbf{Memory Content.}
\textit{A user interested in personal development and psychological insights, particularly through the lens of Buddhist teachings. Enjoys literature that offers wisdom and understanding of the human experience, and is now exploring visual storytelling through unique editions like stereoscopic books.}

\medskip
\textbf{Atomic Fields.}
\begin{itemize}
    \item \textbf{Keywords:} personal development, Buddhist teachings, psychology, wisdom, human experience, visual storytelling, stereoscopic books
    \item \textbf{Tags:} self-help, spirituality, Buddhist, personal growth, visual arts
    \item \textbf{Context:} User seeks literature that provides psychological insights and personal growth, now showing interest in visual storytelling.
\end{itemize}

\medskip
\textbf{Prediction.}
The ground-truth item is \textit{The King of Style: Dressing Michael Jackson}, ranked at position 3. The two higher-ranked items are \textit{Teresa of Avila: The Progress of a Soul} and \textit{Victory in Singleness: A Strategy for Emotional Peace}.

\medskip
\textbf{Diagnosis.}
The model does not make an unreasonable prediction. The final memory strongly emphasizes Buddhist teachings, psychology, personal development, spirituality, and personal growth. It also captures a weaker signal about visual storytelling and stereoscopic books, which suggests that the memory is trying to absorb later visual-format interests. However, the target item about Michael Jackson's style is a sharper shift toward fashion, music culture, and visual celebrity aesthetics. The model therefore follows the dominant long-term preference trajectory rather than this idiosyncratic short-term deviation.

\medskip
\textbf{Takeaway.}
This case exposes a tension between long-term memory consistency and short-term exploration. The evolved memory is stable and coherent, but the ground-truth item reflects a transient preference shift that is weakly supported by the user's prior history.

\end{casebox}

\textbf{Case C: Successful Memory--Target Alignment.} We also include a successful example as a contrast to the previous two cases. Here, the evolved memory center aligns closely with the target item, showing that atomic abstraction and memory evolution can support accurate top-ranked recommendation when the target follows the user's dominant preference trajectory.

\begin{casebox}{Case C: Successful Alignment Between Evolved Memory and Target}

\textbf{User.} User 4991

\medskip
\textbf{Final Memory Snapshot.}
\begin{itemize}
    \item \textbf{Note ID:} \texttt{N-190773}
    \item \textbf{Revision:} 3
    \item \textbf{History Reference:} \texttt{N-190771}
    \item \textbf{Links:} \texttt{N-167292}, \texttt{N-142015}, \texttt{N-177460}, \texttt{N-149729}
\end{itemize}

\medskip
\textbf{Memory Content.}
\textit{User 4991 is interested in personal stories of resilience and transformation, particularly those involving near-death experiences and life reflections. They appreciate narratives that explore profound life changes and the human experience, and are now also drawn to themes of manifestation, personal empowerment, and psychological insights through the lens of Buddhist teachings, including visual storytelling.}

\medskip
\textbf{Atomic Fields.}
\begin{itemize}
    \item \textbf{Keywords:} personal stories, resilience, transformation, near-death experience, life reflections, manifestation, empowerment, psychology, Buddhist teachings, visual storytelling
    \item \textbf{Tags:} biography, self-discovery, personal growth, transformation, spirituality, visual arts
\end{itemize}

\medskip
\textbf{Prediction.}
The ground-truth item is \textit{Driving Straight on Crooked Lines}, ranked at position 1.

\medskip
\textbf{Diagnosis.}
This case shows when memory evolution works well. The target item aligns with the user's evolved memory center, including personal narrative, resilience, transformation, self-discovery, and spiritual reflection. Unlike User 6711, the memory abstraction is not compressed into an overly narrow subtheme. Unlike User 5356, the target does not deviate sharply from the dominant preference trajectory.

\medskip
\textbf{Takeaway.}
When the target item is well aligned with the evolved memory center, atomic memory abstraction and collaborative evolution can support accurate top-ranked recommendation.

\end{casebox}

\begin{casebox}{Additional Observation: Semantic Absorption vs. Link Traceability}

\textbf{Observed Link Statistics.}
\begin{itemize}
    \item \textbf{User 6711:} \texttt{cross\_user\_links = 0}, \texttt{cross\_item\_links = 0}
    \item \textbf{User 5356:} \texttt{cross\_user\_links = 1}, \texttt{cross\_item\_links = 3}
    \item \textbf{User 4991:} \texttt{cross\_user\_links = 0}, \texttt{cross\_item\_links = 4}
\end{itemize}

\medskip
\textbf{Observation.}
For User 6711, the latest memory does not preserve explicit collaborative links, even though earlier traces suggest that collaborative evidence contributes to memory evolution.

\medskip
\textbf{Diagnosis.}
This suggests that collaborative signals may be absorbed into memory text, keywords, tags, and contextual descriptions, but are not always retained as explicit final-step links. In other words, collaborative evidence can shape memory semantics while becoming less traceable in the final memory graph.

\medskip
\textbf{Implication.}
Future work could improve structural fidelity by introducing link persistence, confidence-aware link decay, and provenance tracking during memory evolution.

\end{casebox}

These cases reveal three complementary behaviors. User 4991 shows a successful case where the evolved memory center aligns with the target item, yielding a top-1 hit. User 6711 shows that memory evolution can capture the correct broad preference region but over-compress it into a narrow subtheme, causing semantically adjacent false positives. User 5356 shows that stable long-term memory may under-rank idiosyncratic target items that reflect short-term deviations. Together, these cases suggest that future work should improve fine-grained reranking, preserve explicit collaborative provenance, and better distinguish persistent preference evolution from exploratory behavior.

\section{Additional Limitations and Deployment Considerations}
\label{app:broader_limitations}

\textbf{Privacy and Governance.} \textsc{AtomRec} builds semantic links across user and item memories, which may raise privacy and governance concerns in real deployments. Although our experiments use public benchmark data and do not involve personally identifiable information, fine-grained atomic notes may still encode sensitive preference traces. Practical systems should support data minimization, user-level deletion, provenance tracking, and access control for cross-user links. One possible deployment option is to restrict semantic linking within privacy-preserving user groups or user-siloed settings, where only aggregated or anonymized memory evidence can be shared across users.

\textbf{Latency and Deployment.} Our efficiency analysis focuses on token cost and average memory statistics rather than end-to-end serving latency under concurrent traffic. Real latency depends on API infrastructure, batching, caching, deployment hardware, and whether memory-side operations are executed online or asynchronously. In practice, atomic construction, semantic linking, and memory evolution can be performed after interactions and cached for later recommendation. Future work should evaluate p50/p90 latency and throughput in online serving environments.

\textbf{Backbone Dependence.} We primarily evaluate \textsc{AtomRec} with \texttt{gpt-4o-mini}. Although the framework is model-agnostic in design, different open-source backbones may produce different link, evolution, and reranking quality. This dependence may affect reproducibility for groups without access to the same proprietary model. Evaluating stronger open-source memory and recommendation agents is an important direction for future work.

\section{Extended Related Work}
\label{sec:appendix_related_work}

This section provides a comprehensive review of the literature pertinent to our framework, detailing the evolution of memory mechanisms in Large Language Models (LLMs) and the trajectory of LLM-based autonomous agents in recommender systems.

\textbf{Memory Mechanisms in LLM Agents.} While LLMs excel in long-horizon reasoning \cite{gpt4, gpt-hf, gpt-zeroshot, react,lcz1,lcz2}, their restricted context windows necessitate external memory for persistent knowledge retention \cite{memory-survey1, memory-survey2, memory-survey3, agent1, agent2, lcz3,lcz4}. Early architectures rely on static workflows, such as MemoryBank's forgetting curves \cite{memorybank}, MemGPT's hierarchical buffers \cite{memgpt}, and SCM's read-write streams \cite{scm}. To enhance adaptability, recent systems (e.g., A-Mem \cite{a-mem}, Mem0 \cite{mem0}, MemInsight \cite{meminsight}) transition to dynamic paradigms via associative linking and retrospective summarization \cite{selfevolve, rag-survey}. However, these designs overwhelmingly optimize for isolated, single-agent environments, leaving collaborative memory—where multiple agents co-evolve an interconnected memory space—largely uncharted for interactive recommendation.

\textbf{Large Language Models for Recommendation (LLM4Rec).} Before the advent of autonomous agents, research adapted LLMs for recommendation through various paradigms\cite{lcz5, lcz6, lcz7, lcz8, lcz9, lcz10}. Early approaches leveraged prompting (Chat-REC \cite{chat-rec}, zero-shot rankers \cite{zeroshot}) or instruction tuning and task unification (P5 \cite{p5}, InstructRec \cite{instructRec}, TALLRec \cite{tallrec}) to align LLMs with domain-specific patterns. To resolve vocabulary mismatches, generative models introduced Semantic ID (SID) paradigms (TIGER \cite{tiger, tiger1}, LC-Rec \cite{LC-Rec}, MinioneRec \cite{minionerec}) to natively generate structured identifiers and enhance cross-domain generalization. Concurrently, efforts like Cot4Rec \cite{cot4rec} injected Chain-of-Thought (CoT) reasoning. Despite these advancements, traditional LLM4Rec models remain static, single-turn predictors lacking the autonomy, interactive tool-use, and evolving memory required for dynamic environments, thus catalyzing the shift towards agentic systems.

\textbf{Agents-based Recommendation.} The integration of agentic capabilities into recommender systems has shifted the paradigm from passive modeling to active planning \cite{agent4rec-survey1, agent4rec-survey2, agent4rec-survey3}, predominantly categorized into simulation and recommender-oriented approaches. Simulation frameworks deploy agents as digital twins to model interaction dynamics; for instance, Agent4Rec \cite{agent4rec} uses agents as user simulators, while AgentCF \cite{agentcf} proposes a bidirectional multi-type (user-item) agent simulation. Conversely, recommender-oriented systems design agents to execute complex tasks. While initial efforts focused on empowering a single centralized agent through self-inspired planning (RecMind \cite{recmind}) or brain-toolbox architectures (InteRecAgent \cite{interrec}), recent pioneering works like MACRec \cite{macrec} enable adaptable multi-agent collaboration to handle diverse and complex user intents.

\textbf{Bridging the Gap: Memory in Agentic RS.} To sustain long-term personalization, integrating explicit memory into agentic RS is essential, yet most systems update memories in isolation. For example, iAgent \cite{iagent} and RecBot \cite{recbot} confine updates to individual user profiles via self-reflection, entirely discarding high-order collaborative connectivity. To address this, the state-of-the-art framework MemRec \cite{memrec} constructs a macroscopic collaborative memory graph. However, MemRec fundamentally suffers from coarse-grained, node-level representations that obfuscate specific intents, and rigid graph propagation rules bounded by explicit historical interactions. In contrast, our proposed framework comprehensively deconstructs monolithic memories into multi-attribute \textit{atomic notes} and replaces predefined edges with LLM-driven \textit{autonomous semantic linkage}, achieving dynamic memory evolution at a granular level and providing a structured basis for context-aware collaborative reasoning.

\section*{Generative AI Statement}

Generative AI tools were used only for language polishing and grammar checking. All technical contributions, experiments, analyses, and conclusions were developed and verified by the authors, who take full responsibility for the paper.

\end{document}